# Ground States and Flux Configurations of the Two–dimensional Falicov–Kimball Model

Christian Gruber[1], Nicolas Macris[1], Alain Messager[2] and Daniel Ueltschi[1]

[1] Institut de Physique Théorique
EPFL, CH–1015 Lausanne, Switzerland

[2] Centre de Physique Théorique CNRS
Luminy, Case 907, F–13288 Marseille Cedex 9, France

The Falicov–Kimball model is a lattice model of itinerant spinless fermions ("electrons") interacting by an on–site potential with classical particles ("ions"). We continue the investigations of the crystalline ground states that appear for various filling of electrons and ions, for large coupling. We investigate the model for square as well as triangular lattices. New ground states are found and the effects of a magnetic flux on the structure of the phase diagram is studied. The flux phase problem where one has to find the optimal flux configurations and the nuclei configurations is also solved in some cases. Finaly we consider a model where the fermions are replaced by hard–core bosons. This model also has crystalline ground states. Therefore their existence does not require the Pauli principle, but only the on–site hard–core constraint for the itinerant particles.

# Contents

# 1  Introduction

The occurence of long–range order in itinerant electron models, like the Hubbard model, is usually very difficult to prove. At half–filling on a cubic two or three dimensional lattice the Hubbard model is supposed to display Néel order at low temperature ($d \geq 3$) or in the ground state ($d = 2$). A simpler model where a proof of this fact has been achieved is the so called Falicov–Kimball [12]. There, only one species of electrons (say the spin up electrons) is allowed to hop, while the other type of electrons (spin down electrons) are treated as a classical lattice gas. A nice interpretation of the model is that of a model for matter, where the static particles correspond to heavy ions which interact with the electrons via an on–site potential. The system displays crystalline long range order for the ions which comes from the interplay between the on–site interaction of electrons and ions and the electronic kinetic energy. The Fermi statistics of the electrons plays an important role, because if the electrons are replaced by regular bosons then the ions prefer to clump together [12]. An open question which we solve in this work is what happens if fermions are replaced by hard–core bosons.

On a bipartite lattice and for a special value of the chemical potentials $(\mu_e, \mu_i)$ of the electrons and ions, it was shown in [12] that at low temperatures the ions tend to occupy one of the two sublattices. On a square lattice this corresponds to a chessboard configuration. This value of $(\mu_e, \mu_i)$ is such that the system possesses a particle–hole symmetry. In what follows we fix this special value at $(\mu_e, \mu_i) = (0,0)$ and refer to it as the symmetry point (see paragraph 2.1). This result was then extended to a narrow strip of chemical potentials around the symmetry point, provided the coupling between electrons and nuclei is strong [16].

Concerning the ground states of the model more results are available. The study of the zero temperature phase diagram in the chemical potential plane $(\mu_e, \mu_i)$ was undertook in [7, 8] and for one dimension in [9]. For large coupling, which is the case that will interest us in this paper, formal perturbation expansion shows that away from the symmetry point new periodic ground states appear, with rational densities of ions different from one half. The existence of some of these ground states was confirmed rigorously in [11], where a canonical setting was used. For the fixed densities 1/3, 1/4, 1/5, the ground state configurations are determined rigorously using perturbation theory for large coupling.

In the present work we continue the investigation of the phase diagram in the chemical potential plane $(\mu_e, \mu_i)$ in a more systematic way. Our results generalize and extend what was previously known in three directions : i) we consider square as well as triangular lattices, ii) we allow for Fermi statistics for the electrons as well as hard–core boson statistics, iii) we investigate the effects of a magnetic field which can be homogeneous or inhomogeneous.

Let us briefly describe the new features which appear.

i) *Triangular lattice.* To our knowledge nothing rigorous is known for the Falicov-Kimball model on a triangular lattice. This lattice is not bipartite and in general the

particle–hole symmetry is not present. However it is recovered for a special magnetic field, namely a flux equal to $\pm\frac{\pi}{2}$ through each triangle (see the end of paragraph 2.1 for more details). This fact has been exploited recently in [24] to extend the uniform density theorem [20] to non bipartite lattices with special magnetic fluxes. To leading order of the perturbation theory one sees that the effective interaction between the nuclei is of the Ising type (for any lattice). Thus when the lattice is triangular there appears frustration effects at zero chemical potential $\mu_e = \mu_i = 0$. An interesting question is therefore what happens when the higher order terms of the perturbation theory are included : does this frustration disappears or does it persist ? Our study indicates that frustration disappear when higher terms are included. In fact for zero magnetic field it is rigorously established that there is no frustration at $\mu_e = \mu_i = 0$ for large enough coupling constant.

ii) *Quantum statistics.* The "electrons" will either be regular fermionic particles or hard–core bosons. For usual bosons, as stated before, there can be no ordering for the ions which prefer to clump together. On the contrary, for hard–core bosons there appears crystalline order for a finite range of densities. The ground states may be different from those of fermions. However for the square lattice around the symmetry point the situation is the same as with the Fermi statistics : the ground state configuration of the ions are the chessboard states. Here the methods that we use are restricted to large coupling, but this last result can be proved to hold for all couplings, by using a reflection positivity technique [22].

In the case of Fermi statistics, given a configuration of ions, we can always reduce the problem to a one particle hamiltonian. One can then compute the ground state energy simply by filling the levels (they depend on the configuration of nuclei) of the one particle hamiltonian according to the Pauli principle. For hard–core bosons the situation is completely different a priori. The hard–core introduces a non trivial two body interaction between the bosons. Therefore we have a genuinely many particle hamiltonian. To perform the perturbation expansion we use a functional representation recently developped by S. Miracle–Solé and A. Messager [25] for the model in the fermionic case. In fact for hard–core bosons the method is perfectly well adapted and is even simpler. It turns out that the effective potential between the ions is qualitatively similar to that of the fermionic case, and also leads to crystalline ground states.

The chemical potentials plane is decomposed into domains corresponding to a fixed density of electrons and ions. Therefore the compressibility vanishes at zero temperature and these states are insulating. It is interesting to note that in the bosonic case this behaviour is proven for a truly many particle interacting system.

iii) *Magnetic fields.* The hopping amplitude of the itinerant fermions or bosons can be complex. The sum of the phases around a circuit of the lattice can be interpreted as the flux of an external magnetic field. Our results allow to see how the ground states are affected when one varies the magnetic field. It is found that in general the crystalline configurations of the nuclei can vary with the magnetic field. In fact this is not surprising since in this model the crystalline order is produced by a combination of the potential and the kinetic energies.

Another problem that we address is the so–called flux phase problem. This problem arises in the context of theories for high $T_c$ superconductors [1, 15], but is also important from the point of view of understanding electronic diamagnetism of itinerant electrons on two–dimensional lattices [10]. In the present context one fixes the flux configuration through each face (plaquette or triangle) of the lattice. The problem is to minimize the ground state energy simultaneously with respect to the ion and flux configurations. In the bosonic case it is easy to see that the optimal flux is always zero (diamagnetic inequality). For fermions however the situation is more subtle. We find that for large coupling, the optimal flux configuration consists of zero and $\pi$ through the faces of the lattice. On the square lattice, for density $\frac{1}{2}$ the optimal flux configuration is $\pi$ through each plaquette; for density $\frac{1}{3}$ the optimal flux configuration is not uniform, but of period three, the average flux being $\frac{\pi}{3}$; for density $\frac{1}{5}$ one finds that the flux is 0 for sufficiently large coupling. We refer the reader to section 4 for more details and the results on a triangular lattice. It has been conjectured in the literature that for free fermions on a square lattice the optimal flux should be equal to $2\pi n$, where $0 \leq n \leq 1$ is the filling factor [10]. Our results cannot however be compared with this conjecture since we work at large coupling, and we are therefore far from the limit of free electrons. On the other hand they show that the optimal flux configuration is not necessarily uniform as it is often assumed in the literature. Recently the conjecture has been proven for the half–filled band of the Hubbard model [18] (see also [23]).

The paper is organized as follows. In section 2 we derive the $U^{-1}$ expansions for the fermionic and bosonic cases for square and triangular lattices. The phase diagrams for a uniform magnetic flux are then studied in section 3. Section 4 discusses the optimal flux configurations of the model. The appendices contain more technical material.

## 2   $U^{-1}$ – expansion for the Falicov–Kimball Models

### 2.1   Definition of the models

The spinless Falicov-Kimball models have been introduced to discuss lattice systems consisting of identical quantum particles interacting with classical particles (or "ions") by means of on-site interaction which mimics the Coulomb repulsion or attraction. The quantum particles can be subjected to an external magnetic field which appears in their kinetic energy. We shall investigate several models in which the quantum particles can be either fermions or hard-core bosons; the lattice on which the systems are defined is either a square or a triangular lattice.

We shall consider only finite lattices $\Lambda$ and for technical reasons we introduce periodic boundary conditions. The quantum particles are represented by creation and annihilation operators, which satisfy the usual "commutation" relations :

$$\text{fermions :} \qquad \{a_x^+, a_y^+\} = \{a_x, a_y\} = 0 \; \forall x, y \in \Lambda$$
$$\{a_x, a_y^+\} = \delta_{xy} \tag{2.1}$$
$$\text{hard-core bosons :} \qquad [a_x^\#, a_y^\#] = 0 \; \forall x \neq y \in \Lambda$$
$$(a_x^+)^2 = (a_x)^2 = 0 \qquad \{a_x, a_x^+\} = 1 \tag{2.2}$$

The "ions" are described by a random variable $s_x \in \{-1, +1\}$ such that $s_x = +1$ if the site $x \in \Lambda$ is occupied and $s_x = -1$ if the site $x$ is empty. The finite volume hamiltonian of the Falicov–Kimball models is written as

$$H_\Lambda(s) = - \sum_{x,y \in \Lambda} t_{xy} a_x^+ a_y + U \sum_{x \in \Lambda} s_x (a_x^+ a_x - \frac{1}{2}) \tag{2.3}$$

where $s = \{s_x\}_{x \in \Lambda}$ denotes the configuration of ions, $t_{xy} = |t_{xy}| e^{i\theta_{xy}} = t_{yx}^*$, i.e. $\theta_{xy} = -\theta_{yx}$, and $\theta_{xy} \in [0, 2\pi[$ is related to the integral of the vector potential $\theta_{xy} = \int_x^y \vec{A} \cdot d\vec{x}$. For the cases under investigation $|t_{xy}| = t$ for nearest neighbours and zero otherwise.

Given a circuit $\omega$ of length $|\omega| = n$, i.e. an ordered sequence of sites $(x_1, \ldots, x_n, x_1)$, with $t_{x_j x_{j+1}} \neq 0$ for $j = 1, \ldots, n$, defined up to a cyclic permutation, we introduce the flux across the circuit as

$$\phi_\omega = \sum_{j=1}^n \theta_{x_j x_{j+1}} \;(\text{mod } 2\pi). \tag{2.4}$$

Moreover, we also need to introduce the number $N(\omega)$ of periodic sequences in $\omega$, i.e.

$$N(\omega) = \frac{n}{n'} \tag{2.5}$$

where $n'$ is the length of the shortest circuit $\omega'$ such that $\omega$ is obtained by repeating $N(\omega)$-times the circuit $\omega'$.

If the coupling $U$ is positive the interaction between quantum particles and ions is repulsive, if $U$ is negative it is attractive.

It is important to stress that contrary to the case of fermions where, except for the statistics, we have a system of non-interacting particles under the influence of the external potential $\{U s_x\}$ and the magnetic field, in the case of bosons we have a system of interacting particles. Therefore in the case of fermions one is led to consider the eigenvalues of the one particle hamiltonian, but in the case of bosons we are forced to consider the $N$ particle hamiltonian.

As usual, the *effective interaction* $F_\Lambda$ between the ions is defined by the trace of the Boltzmann factor over the Fock space of the quantum particles

$$e^{-\beta F_\Lambda(\beta, s; \mu_e, \mu_i)} = \text{Tr}\{e^{-\beta[H_\Lambda(s) - \mu_e N_e - \mu_i N_i(s)]}\} \tag{2.6}$$

where

$$N_e = \sum_{x \in \Lambda} a_x^+ a_x \qquad N_i(s) = \frac{1}{2} \sum_{x \in \Lambda} (s_x + 1) \tag{2.7}$$

are the electron and ion numbers, and $\mu_e, \mu_i$ their respective chemical potentials. The zero–temperature limit defines the *effective hamiltonian for the ions*

$$H_{eff}(s; \mu_e, \mu_i) = \lim_{\beta \to \infty} F_\Lambda(\beta; s; \mu_e, \mu_i). \tag{2.8}$$

In the case of fermions the particles do not interact and the hamiltonian (2.3) is the second–quantized version of the one–body hamiltonian (up to the constant term)

$$h_\Lambda(s) = T + US \tag{2.9}$$

$$(T)_{xy} = -t_{xy} \qquad (S)_{xy} = s_x \delta_{xy} \tag{2.10}$$

therefore the effective hamiltonian (2.8) for the ions is the sum of the lowest eigenvalues $e_j(s)$ of $h_\Lambda(s)$

$$\begin{aligned} H_{eff}(s; \mu_e, \mu_i) &= \sum_{e_j(s) \leq \mu_e} (e_j(s) - \mu_e) - \frac{U}{2} \mathrm{Tr} S - \mu_i N_i(s) \\ &= E(s; \mu_e) - \mu_e N_e - \mu_i N_i(s) \end{aligned} \tag{2.11}$$

with

$$E(s; \mu_e) = \sum_{e_j(s) \leq \mu_e} e_j(s) - \frac{U}{2} \mathrm{Tr} S, \qquad N_e = \sum_{e_j(s) \leq \mu_e} 1. \tag{2.12}$$

Let $\Phi = \{\phi_C\}$ denote the flux configuration across elementary cells $C$ (i.e. the plaquettes $P$ or triangles $\Delta$ of $\Lambda$). When $\phi_C = \phi$ for all $C$ we say that $\Phi$ is homogeneous. The problems we want to investigate are the following.

i) given $(\mu_e, \mu_i, \Phi)$, with $\Phi$ homogeneous, find the configurations of ions $\mathcal{S} = \{s^\alpha\}$ which minimize the effective hamiltonian : this is the problem of the ground state, with constant external magnetic field.

ii) given $(\mu_e, \mu_i)$, find the configurations of ions $\mathcal{S}$ and the fluxes $\{\phi_C\}$, not necessarily uniform, which minimize the effective hamiltonian : this is the flux phase problem.

We shall use the $U^{-1}$ expansion of the effective hamiltonian $H_{eff}$ and we shall restrict our considerations to the case

$$|U| > zt \qquad \mu_e \in \, ]-|U|+zt, |U|-zt\,[\,, \qquad t = \sup |t_{xy}| \tag{2.13}$$

where $z$ is the maximal coordination number ($z = 4$ for the square lattice, $z = 6$ for the triangular lattice). This is the domain where the expansion is absolutely and uniformly convergent with respect to $\Lambda$.

Moreover, if $U > 0$ (repulsion), (2.13) implies the "*half-filling condition*" $N_e + N_i(s) = |\Lambda|$ and if $U < 0$ (attraction), (2.13) implies the "*neutrality condition*" $N_e = N_i(s)$.

This is a direct consequence of the properties of the spectrum of the one–body hamiltonian $h_\Lambda(s)$ (for fermions) and also follows from the closed loop expansion (section 2.3) valid for fermions and bosons.

Finally we want to stress the role played by the statistics and the nature of the lattice on the symmetries of the phase diagrams. First of all if $s = \{s_x\}_{x \in \Lambda}$ is a ground state for $U$ and $(\mu_e, \mu_i)$, then $\bar{s} = \{-s_x\}_{x \in \Lambda}$ is a ground state for $-U$ and $(\mu_e, -\mu_i)$. This is true for any lattice, any statistics and any flux configuration. Therefore we can restrict our discussion to the case $U > 0$ in the next sections.

For bipartite lattices and fermions (the square lattice for example) the phase diagram has the following symmetry. If $s$ is a ground state for some value $(\mu_e, \mu_i)$ and some flux configuration $\{\phi_P\}$, then $\bar{s}$ is a ground state for $(-\mu_e, -\mu_i)$ and $\{-\phi_P\}$. This follows from the unitary particle–hole transformation $a_x^+ \to a_x$, $a_x \to a_x^+$ applied to the hamiltonian (2.3). By a time reversal transformation (complex conjugation) we see that $\bar{s}$ is also a ground state for $(-\mu_e, -\mu_i)$ and $\{\phi_P\}$. In particular this implies that figures 7 and 15 in sections 3 and 4 are symmetric with respect to the line $\mu = 0$.

For the triangular lattice and fermions (non bipartite case) the above symmetry of the phase diagram is in general lost. However the following holds. If $s$ is a ground state for some value $(\mu_e, \mu_i)$ and some flux configuration $\{\phi_\Delta\}$, then $\bar{s}$ is a ground state for $(-\mu_e, -\mu_i)$ and $\{\pi - \phi_\Delta\}$. This follows again from the same particle–hole transformation. By time reversal the statement still holds with $\{\pi - \phi_\Delta\}$ replaced by $\{\pi + \phi_\Delta\}$. In particular if $\phi_\Delta = \pm\frac{\pi}{2}$ through all triangles then $\pi - \phi_\Delta = \pm\frac{\pi}{2}$, and we have a symmetry of the hamiltonian as mentioned in the introduction. As a consequence on the figures 10, 11, 12 the horizontal line $\phi = \frac{\pi}{2}$ is symmetric with respect to the vertical line $\mu = 0$. In figure 16 the case $\mu > 0$ can be obtained from $\mu < 0$ by interchanging the values 0 and $\pi$ for the fluxes and replacing $s$ by $\bar{s}$.

In the case of hard–core bosons the particle–hole transformation $a_x^+ \to a_x$, $a_x \to a_x^+$ leads to the following property, irrespective of the type of lattice. If $s$ is a ground state for $(\mu_e, \mu_i)$ and $\{\phi_C\}$ then $\bar{s}$ is a ground state for $(-\mu_e, -\mu_i)$ and $\{-\phi_C\}$. By time reversal $\bar{s}$ is also a ground state for $(-\mu_e, -\mu_i)$ and $\{\phi_C\}$. Therefore the phase diagrams of figures 8 and 13 are symmetric around the line $\mu = 0$.

## 2.2   $U^{-1}$ expansion of $H_{eff}$ for fermions

For fermions systems satisfying the condition (2.13), we can use the representation

$$E(s; \mu_e) = \frac{1}{2\pi i} \oint_{\mathcal{D}} dz \, \mathrm{Tr}\left(\frac{z}{z - h_\Lambda(s)}\right) - \frac{U}{2}\mathrm{Tr}S_\Lambda \tag{2.14}$$

where $\mathcal{D}$ is a circle in the complex $z$-plane, centered at $-|U|$ and enclosing all negative eigenvalues. By expanding the resolvent we obtain the following absolutely convergent expansion in powers of $U^{-1}$

$$H_{eff}(s; \mu_e, \mu_i) = -\mu_i N_i(s) - \mu_e N_e - \frac{|U|}{2}|\Lambda|$$

$$+\sum_{k\geq 2}\frac{1}{(2|U|)^{k-1}}\sum_{\omega\in\Omega_k(s)}(-1)^{m(\omega)}t_{x_1x_2}\ldots t_{x_kx_1}\frac{(k-2)!}{(m(\omega)-1)!(k-m(\omega)-1)!}\frac{1}{N(\omega)} \quad (2.15)$$

where

$$N_e = \begin{cases} |\Lambda| - N_i(s) & \text{if } U > zt \\ N_i(s) & \text{if } U < -zt. \end{cases}$$

In the above formula $\Omega_k(s) = \{\omega \in \Omega(s); |\omega| = k\}$, where $\Omega(s)$ is the set of circuits $\omega = (x_1, \ldots, x_k, x_1)$ which contain at least one empty site and one occupied site. For $U > 0$, $m(\omega)$ is the number of sites in $(x_1, \ldots, x_k)$ with $s_{x_j} = -1$, i.e. the number of empty sites in $\omega$; for $U < 0$, $m(\omega)$ is the number of occupied sites in $(x_1, \ldots, x_k)$.

In the particular case where $|t_{xy}| = t$ if $(x,y)$ are nearest neighbours and zero otherwise (2.15) yields

$$H_{eff}(s; \mu_e, \mu_i) = -\mu_i N_i(s) - \mu_e N_e - \frac{|U|}{2}|\Lambda|$$
$$+ \sum_{k\geq 2}\frac{1}{|U|^{k-1}}\sum_{\omega\in\Omega_k(s)} g_\omega(s)\cos\phi_\omega(s) \quad (2.16)$$

where

$$g_\omega(s) = (-1)^{m(\omega)}\frac{1}{2^{k-1}}\frac{(k-2)!}{(m(\omega)-1)!(k-m(\omega)-1)!}\frac{1}{N(\omega)} \quad (2.17)$$

To conclude this discussion let us note that for bipartite lattices the sum in Eq. (2.15) is only over even values of $k$.

## 2.3 Closed loop expansion for $F$

For hard–core bosons a similar $U^{-1}$ expansion is valid, although its derivation is very different. We use a method first invented (for fermions) by A. Messager and S. Miracle-Sole [25]. The basic steps are illustrated here for the convenience of the reader, in the case $U > 0$.

Thanks to the Trotter product formula

$$e^{-\beta(H_0+H_1)} = \lim_{M\to\infty}\left[(1-\frac{\beta}{M}H_0)e^{-\frac{\beta}{M}H_1}\right]^M \quad (2.18)$$

with $H_0$ the kinetic part of (2.3) and $H_1$ the interaction part. We can express the trace over the Fock space of the quantum particles in the form

$$e^{-\beta F(\beta;s;\mu_e,\mu_i)} = \sum_{N_e\geq 0} e^{\beta(\mu_i N_i + \mu_e N_e)}e^{\beta U(N_i+N_e-\frac{|\Lambda|}{2})}$$
$$\cdot \lim_{M\to\infty}\sum_{X_0}\ldots\sum_{X_{M-1}}\varepsilon e^{-2\frac{\beta}{M}U\sum_{i=1}^{M-1}v(X_i)}\prod_{i=1}^M a(X_i; X_{i-1}) \quad (2.19)$$

where

$$X = \{x_1, x_2, \ldots, x_{N_e}\} = \text{ordered subset of } \Lambda$$
$$v(X) = |\{x \in X; s_x = 1\}| = \text{number of doubly occupied sites}$$
$$\text{(by one ion and one electron)}$$
$$X_M = X_0 \text{ up to permutations}$$

and

$$a(Y; X) = \begin{cases} 1 & \text{if } Y = X \\ \frac{\beta}{M} t_{yx} & \text{if } Y \text{ differs from } X \text{ by one jump from } x \text{ to } y \\ 0 & \text{otherwise} \end{cases}$$

$\varepsilon = +1$ for hard-core bosons

$\varepsilon = $ sign of the permutation $X_0 \to X_M$ for fermions.

Introducing the $(1+\nu)$-dimensional lattice, where the new dimension is related to the "time" variable $x_n^0 = \frac{\beta}{M} n$, $n = 0, \ldots, M-1$, and all the functions defined on this lattice are periodic in $x^0$ with period $\beta$, one is led to consider the family of oriented closed loops on the cylinder $[0, M-1] \times \Lambda$ (not to be confused with the circuits defined previously). The following expansion can be obtained for $|t_{xy}| = t$

$$F(\beta; s; \mu_e, \mu_i) = (\mu_e - \mu_i) N_i - (\mu_e + \frac{U}{2})|\Lambda| + F'(\beta; s; \mu_e) \tag{2.20}$$

$$F'(\beta; s; \mu_e) = -\frac{1}{\beta} \lim_{M \to \infty} \log \left\{ \sum_{\underline{\omega} \sim s}^{*} \prod_{\omega_j \in \underline{\omega}} K(\omega_j) \right\} \tag{2.21}$$

$$K(\omega_j) = e^{-\frac{\beta}{M}(U - \mu_e)|\omega_{j\uparrow}|} e^{-\frac{\beta}{M}(U + \mu_e)|\omega_{j\downarrow}|} \left(\frac{\beta t}{M}\right)^{h(\omega_j)} e^{i\phi(\omega_j)} \varepsilon(\omega_j) \tag{2.22}$$

The sum $\sum^*$ in (2.21) indicates the sum over non intersecting set of closed loops $\underline{\omega} = (\omega_1, \ldots, \omega_r)$ on the $(1+\nu)$-dimensional lattice with at most one jump at each time $x_n^0$. Moreover $\underline{\omega} \sim s$ means that the closed loops $(\omega_1, \ldots, \omega_r)$ are compatible with $s$, i.e. the orientation of the segments $[(x^0, x), (x^0 + \frac{\beta}{M}, x)]$ is in the positive time direction if $s_x = +1$, in the negative time direction if $s_x = -1$; $|\omega_{j\uparrow}|$ and $|\omega_{j\downarrow}|$ denote the number of segments in the time direction, which are oriented in the positive and in the negative time direction : $\sum_j [|\omega_{j\uparrow}| - |\omega_{j\downarrow}|] = |\Lambda| - N_e - N_i$, $h(\omega_j)$ is the number of horizontal bonds (or jumps) in $\omega_j$; $\phi(\omega) = \sum_j \theta_{x_j x_{j+1}}$ is the flux through the circuit $(x_1, \ldots, x_k)$, projection of $\omega$ on $\Lambda$.

We recognize in (2.21) the partition function of a polymer system in $(1+\nu)$ dimensions. The polymers are the loops $\omega_j$ and they cannot intersect.

Using the standard cluster expansion we obtain the $U^{-1}$ expansion of the effective interaction $F$ in terms of connected closed loops

$$-\beta F'(\beta; s; \mu_e) = \lim_{M \to \infty} \sum_{r=1}^{\infty} \frac{1}{r!} \sum_{\omega_1, \ldots, \omega_r} K^T(\omega_1, \ldots, \omega_r) \tag{2.23}$$

where $K^T$ is the truncated function. It has been shown in [25] that the above cluster expansion is absolutely convergent, uniformly with respect to $M$ and $\beta$, if $U > 4zt$ and $|\mu_e| < U - 4zt$.

**Remarks**

1. In one dimension $\varepsilon(\omega) = 1$ for all $\omega$ and therefore we recover the known result that hard–core bosons are identical to fermions in one dimension (with open boundary conditions).

2. In the case of fermions the condition of convergence $|\mu_e| < U - zt$, implies that $\mu_e$ is in the gap separating positive and negative eigenvalues of $h_\Lambda(s)$.

3. In the zero temperature limit, and $|\mu_e| < U - 4zt$, (2.21) implies

$$H_{eff}(s;\mu_e,\mu_i) = -(\mu_i - \mu_e)N_i(s) - (\mu_e + \frac{U}{2})|\Lambda|$$
$$- \lim_{\beta \to \infty} \frac{1}{\beta} \left[ \lim_{M \to \infty} \sum_{r=1}^{M} \frac{1}{r!} \sum_{\omega_1,\ldots,\omega_r} K^T(\omega_1,\ldots,\omega_r) \right] \quad (2.24)$$

and only loops which do not go around the cylinder $[0, M-1] \times \Lambda$ have to be considered. Indeed for any loop going around the cylinder there is a factor $\frac{1}{\beta}e^{-\beta U}$ which gives zero in the limit $\beta \to \infty$.

Therefore in the limit $\beta \to \infty$, for each closed loop we have $|\omega_{j\uparrow}| = |\omega_{j\downarrow}|$, and $N_e + N_i(s) = |\Lambda|$ which is the half-filling condition (as it is expected since $\mu_e$ is in the gap).

## 2.4 Flux phase problem for bosons

In (2.23), the right hand side is increased if each term of the sum is replaced by its modulus. This amounts to set $\varepsilon = +1$ and to replace $t_{xy}$ by $|t_{xy}|$, i.e. to set $\theta_{xy} = 0$. Therefore $\exp(-\beta F')$ is increased if we replace the flux by $\phi_C = 0$. The effective interaction, and as a consequence the ground state energy, are minimized if all the fluxes are zero, for all values of $(\mu_e, \mu_i, U, \beta)$. This property of bosons is in fact very general (see for example [26]). For fermions the situation is more subtle and is the subject of section 4.

## 2.5 Expansion up to order 3 for fermions systems

In the case of fermions systems, using Eq. (2.15) one obtains immediately the first few terms of the effective hamiltonian of ions, for $U$ positive. Here the flux configuration is uniform.

**Square lattice**

$$H_{eff}(s;\mu_e,\mu_i) = -\frac{1}{2}(\mu_i - \mu_e)\sum_{x \in \Lambda} s_x - \frac{1}{2}(\mu_e + \mu_i + U)|\Lambda|$$
$$+ \sum_{<x,y>\subset\Lambda} \left[ \frac{t^2}{4U} - \frac{t^4}{16U^3}(7 + 2\cos\phi) \right] s_x s_y + \sum_{\substack{(x,y)\subset\Lambda \\ |x-y|=\sqrt{2}}} \frac{t^4}{16U^3}(4 - \cos\phi)s_x s_y$$

$$+ \sum_{\substack{(x,y) \subset \Lambda \\ |x-y|=2}} \frac{t^4}{8U^3} s_x s_y + \sum_{P \subset \Lambda} \frac{t^4}{16U^3} (\cos\phi)(1 + 5s_P) + O\left(\frac{t^6}{U^5}\right) \tag{2.25}$$

where the last sum runs over the plaquettes $P$ consisting of four sites, and $s_P$ is the product of the four spins around the plaquette.

**Triangular lattice**

$$H_{eff}(s; \mu_e, \mu_i) = -\frac{1}{2}(\mu_i - \mu_e + \frac{3}{2}\frac{t^3}{U^2}\cos\phi) \sum_{x \in \Lambda} s_x - \frac{1}{2}(\mu_e + \mu_i + U)|\Lambda|$$
$$+ \sum_{<x,y> \subset \Lambda} \left[\frac{t^2}{4U} - \frac{t^4}{16U^3}(7 + 5\cos 2\phi)\right] s_x s_y + \sum_{\substack{(x,y) \subset \Lambda \\ |x-y|=\sqrt{3}}} \frac{t^4}{16U^3}(4 - \cos 2\phi) s_x s_y$$
$$+ \sum_{\substack{(x,y) \subset \Lambda \\ |x-y|=2}} \frac{t^4}{8U^3} s_x s_y + \sum_{\Delta \subset \Lambda} \frac{3}{8}\frac{t^3}{U^2}(\cos\phi) s_\Delta$$
$$+ \sum_{P \subset \Lambda} \frac{t^4}{16U^3}(\cos 2\phi)(1 + 5s_P) + O\left(\frac{t^5}{U^4}\right) \tag{2.26}$$

with $\Delta$ a triangle and $P$ two triangles sharing an edge; $s_\Delta$, $s_P$ are the product of their respective spins.

## 2.6 Expansion up to order 3 for bosons systems

For bosons systems we have to use the closed loop expansion (2.24). Let us first note that using the remark 3 of Sec. 2.3, the activity (2.22) of the loop $\omega$ is

$$K(\omega) = e^{-2\frac{\beta}{N}U|\omega_\uparrow|} \left(\frac{\beta t}{N}\right)^{h(\omega)} e^{i\phi[\omega]} \varepsilon(\omega) \tag{2.27}$$

We then observe from Eq. (2.23) that the contribution in $U^{-k}$ will come from loops with $(k+1)$ horizontal jumps (i.e. the coefficient of $t^{k+1}$).

Let us denote by $(x_1, \ldots, x_k)$ the circuit defined by the projection (along the time direction) of the loops $(\omega_1, \ldots, \omega_r)$ on $\Lambda$. In this manner the sum over the loops with $(k+1)$ jumps can be expressed as a sum over circuits. Now if for a given circuit $(x_1, \ldots, x_k)$ all the possible families of closed loops $(\omega_1, \ldots, \omega_r)$ associated with this circuit have the same factor $\varepsilon(\underline{\omega}) = \prod_{j=1}^{r} \varepsilon(\omega_j)$, then the contribution of those loops for bosons systems is immediately obtained from those already obtained for fermions systems.

In particular if the circuit $(x_1, \ldots, x_k)$ is "linear" (does not enclose a plaquette), then one can check that $\varepsilon(\underline{\omega}) = 1$. Furthermore if the circuit is self–avoiding ($x_j \neq x_k$ for all $j \neq k$), then there is exactly one loop and in this case $\varepsilon(\omega) = \varepsilon^{m-1}$ ($\varepsilon = +1$ for bosons, $\varepsilon = -1$ for fermions). Therefore, in this case $\varepsilon(\omega) = 1$ for bosons and $\varepsilon(\omega) = -s_\omega$ for fermions.

These situations will hold for the expansion up to order 3 and therefore we obtain immediately the following expressions for bosons systems. One should note that the above simple properties which enable us to obtain directly the expansion for bosons from the corresponding expression for fermions do not work any more already at order 4 and one has then to use Eq. (2.24).

**Square lattice**

$$H_{eff}(s; \mu_e, \mu_i) = -\frac{1}{2}(\mu_i - \mu_e) \sum_{x \in \Lambda} s_x - \frac{1}{2}(\mu_e + \mu_i + U)|\Lambda|$$
$$+ \sum_{<x,y> \subset \Lambda} \left[ \frac{t^2}{4U} - \frac{t^4}{16U^3}(7 - 2\cos\phi) \right] s_x s_y + \sum_{\substack{(x,y) \subset \Lambda \\ |x-y|=\sqrt{2}}} \frac{t^4}{16U^3}(4 + \cos\phi)s_x s_y$$
$$+ \sum_{\substack{(x,y) \subset \Lambda \\ |x-y|=2}} \frac{t^4}{8U^3} s_x s_y - \sum_{P \subset \Lambda} \frac{t^4}{16U^3}(\cos\phi)(5 + s_P) + O\left(\frac{t^6}{U^5}\right) \qquad (2.28)$$

**Triangular lattice**

$$H_{eff}(s; \mu_e, \mu_i) = -\frac{1}{2}(\mu_i - \mu_e) \sum_{x \in \Lambda} s_x - \frac{1}{2}(\mu_e + \mu_i + U + \frac{3}{2}\frac{t^3}{U^2})|\Lambda|$$
$$+ \sum_{<x,y> \subset \Lambda} \left[ \frac{t^2}{4U} + \frac{t^3}{4U^2}\cos\phi - \frac{t^4}{16U^3}(7 - 5\cos 2\phi) \right] s_x s_y$$
$$+ \sum_{\substack{(x,y) \subset \Lambda \\ |x-y|=\sqrt{3}}} \frac{t^4}{16U^3}(4 + \cos 2\phi)s_x s_y + \sum_{\substack{(x,y) \subset \Lambda \\ |x-y|=2}} \frac{t^4}{8U^3} s_x s_y$$
$$- \sum_{P \subset \Lambda} \frac{t^4}{16U^3}(\cos 2\phi)(5 + s_P) + O\left(\frac{t^5}{U^4}\right) \qquad (2.29)$$

# 3 Phase diagram

## 3.1 Method

In this section we investigate the zero temperature phase diagram, i.e. the ground states of the effective hamiltonian, assuming a constant flux across each plaquette or triangle. In the next section we shall then solve the flux problem, namely given $(\mu_e, \mu_i)$ we shall find the set of fluxes, and the configurations, corresponding to the minimum energy.

We restrict our considerations to the case $U > 4zt$ and $|\mu_e| < U - 4zt$ for which the previous formulæ are valid. In this case the system satisfies the half-filling condition $N_i + N_e = |\Lambda|$, the $U^{-1}$ expansion converges, and the effective hamiltonian is a function of $(\mu_i - \mu_e)$ only. We then take $\mu_e = 0$ and write $\mu_i = \mu$ (which is to say $\mu = \mu_i - \mu_e$).

The idea is to decompose $H_{eff}$ at order $k$ into two parts, the truncated hamiltonian $H^{(k)}$ (which depends on $\mu$), and the rest $R^{(k)}$ (which is independent of $\mu$), i.e.

$$H_{eff} = H^{(k)} + R^{(k)},$$
$$H^{(k)} = -\frac{\mu}{2}\sum_x (s_x + 1) + \sum_{q\geq 1}^{k} \frac{t^{q+1}}{U^q} c_q, \qquad R^{(k)} = \sum_{q=k+1}^{\infty} \frac{t^{q+1}}{U^q} c_q \qquad (3.1)$$

In (3.1) we have dropped the constant $-\frac{1}{2}U|\Lambda|$. Following [11] the truncated part can be written as $H^{(k)} = \sum_B H_B = \sum_B H'_B$ where $H_B$ is given by the $U^{-1}$ expansion, $B \subset \Lambda$ with diam $B \leq \frac{1}{2}(k+1)$, and $H'_B = H_B + K_B$. The "zero–potentials" $K_B$ satisfy

$$\sum_B K_B(s) = 0 \quad \text{for all } s \qquad (3.2)$$

and have the symmetries of the lattice.

We shall then show that there exist $\mu_1 < \mu_2 < \ldots < \mu_n$ ($n$ and $\mu_j$ depending on the order $k$ and the flux configuration $\Phi$) such that for the following results hold :
for $\mu \in ]\mu_j, \mu_{j+1}[$, $j = 0, \ldots, n$, $\mu_0 = -\infty$, $\mu_{n+1} = +\infty$,

a) it is possible to find $K_B$ (depending on $\mu$ and $\Phi$) such that $H'_B$ is an $m$–potential, i.e. there exists a set of periodic configurations $\mathcal{S}_j = \{s_j^\alpha\}$, where the $s_j^\alpha$ are related by the symmetries of the lattice, satisfying the following conditions

$$\begin{cases} H'_B(s) = H'_B(s_j^\alpha) = \min_{\bar{s}} H'_B(\bar{s}) & \text{if } s|_B = s_j^\alpha|_B \\ H'_B(s) > H'_B(s_j^\alpha) & \text{if } s|_B \neq s_j^\alpha|_B \end{cases} \qquad (3.3)$$

Therefore if $\mu \in ]\mu_j, \mu_{j+1}[$, the set of configurations $\mathcal{S}_j$ is the set of ground states for the truncated hamiltonian $H^{(k)}$. On the other hand, for $\mu = \mu_j$ the number of ground states for $H^{(k)}$ is infinite.

b) it is possible to find $\varepsilon_j$, $\delta_j$ positive, $\varepsilon_j + \delta_j < \mu_{j+1} - \mu_j$, such that for $\mu \in ]\mu_j + \varepsilon_j, \mu_{j+1} - \delta_j[$, $\mathcal{S}_j$ is the set of ground states for the effective hamiltonian $H_{eff}$. To establish this result, we follow the ideas introduced in [11] and write the rest as

$$R^{(k)} = \sum_x R_x^{(k)} = \sum_x \bar{R}_x^{(k)} \qquad (3.4)$$

where

$$R_x^{(k)} = \sum_{q \geq k+1} \frac{t^{q+1}}{(2U)^q} \sum_{\substack{\omega \in \Omega_{q+1}(s) \\ \omega \ni x}} (-1)^{m(\omega)} \frac{(q-1)!}{(m(\omega)-1)!(q-m(\omega))!} \frac{1}{N(\omega)|\text{supp } \omega|} \quad (3.5)$$

and $\bar{R}_x^{(k)}$ is the average value of $R_x^{(k)}$ over a cell $C_p(x)$, centered at $x$, associated with the translation symmetry group of the $s_j^\alpha$ (the precise definition of $C_p(x)$ is given later), i.e.

$$\bar{R}_x^{(k)} = \frac{1}{|C_p(x)|} \sum_{y \in C_p(x)} R_y^{(k)}. \quad (3.6)$$

It is then easy to see that

$$\bar{R}_x^{(k)}(s_j^\beta) = \bar{R}_x^{(k)}(s_j^\alpha), \text{ for all } s_j^\beta \in \mathcal{S}_j. \quad (3.7)$$

c) the $\mu_j$, $j = 1, 2, \ldots, n$, are functions of $U$ and $k$ of the form

$$\mu_j = \sum_{q=0}^{k} \frac{t^{q+1}}{U^q} d_q \quad (3.8)$$

and $\varepsilon_j$, $\delta_j$ are of the order $t^{k+2}/U^{k+1}$.

To illustrate the method we consider the order 0 :

$$H_{eff} = -\frac{\mu}{2} \sum_x (s_x + 1) + \sum_{q \geq 1} \frac{t^{q+1}}{U^q} c_q = H^{(0)} + R^{(0)} \quad (3.9)$$

The ground states of $H^{(0)}$ are evidently

$\mathcal{S}_- = \{s_-\}$, where $s_- = \{s_x = -1 \; \forall x\}$ for $\mu < 0$

$\mathcal{S}_+ = \{s_+\}$, where $s_+ = \{s_x = +1 \; \forall x\}$ for $\mu > 0$

For the domain $\mu \leq 0$, let $\mu = -\delta$ and $\Gamma(s) = \{x : s_x = +1\}$,

$$H_{eff}(s) - H_{eff}(s_-) = \delta|\Gamma(s)| + \sum_{q \geq 1} \frac{t^{q+1}}{U^q} [c_q(s) - c_q(s_-)]$$

$$\geq |\Gamma(s)| \left[ \delta - \sum_{q \geq 1} t \left(\frac{zt}{U}\right)^q \right] \quad (3.10)$$

Therefore for $\mu < -z\frac{t^2}{U}(1 - z\frac{t}{U})^{-1}$, $s_-$ is the only ground state for $H_{eff}$. Similarly for $\mu > z\frac{t^2}{U}(1 - z\frac{t}{U})^{-1}$, $s_+$ is the only ground state for $H_{eff}$.

## 3.2 Fermions on the square lattice ($U > 4t$, $|\mu_e| < U - 4t$).

Introducing the next order in the truncated hamiltonian, we obtain the antiferromagnetic Ising model, for which $\mu$ plays the role of the magnetic field (see (2.25)). The phase diagram of the truncated hamiltonian $H^{(1)}$ consists of three distinct regions, corresponding to the empty state, the two chessboard states, and the full state. As before it is possible to bound the contribution of the higher orders and thus to obtain part of the phase diagram of the effective hamiltonian, see fig. 1. The two separation intervals (in thick lines) between the domains have length proportional to $\frac{t^4}{U^3}$ and make place for domains corresponding to ground states with intermediate densities and higher periods.

Let us remark that at this order the phase diagram is independent of the magnetic fluxes.

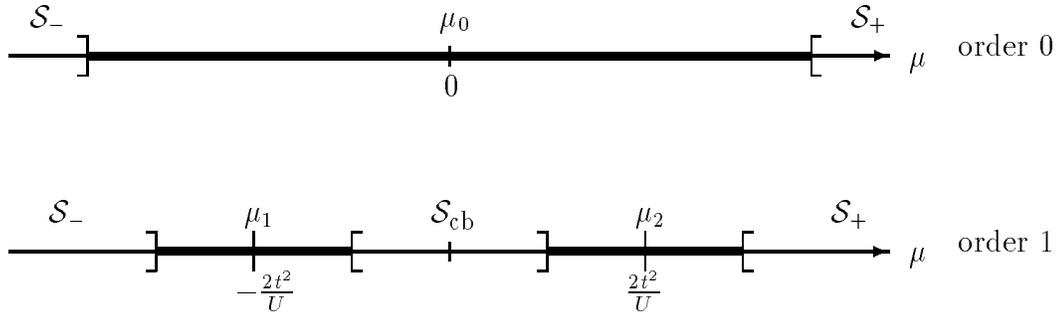

Figure 1: Fermions on a square lattice. Phase diagram obtained at order $U^0$ and $U^{-1}$. The dark domains represent values of $\mu$ where no conclusion is obtained at this order.

Since on the square lattice the $U^{-1}$ expansion of $H_{eff}$ has only odd powers of $U^{-1}$, the next order is $U^{-3}$. Furthermore using the symmetry property of the phase diagram we restrict the discussion to the case $\mu \leq 0$.

*a) Study of $H^{(3)}$.*

We first write the truncated hamiltonian $H^{(3)}$ as a sum over potentials defined on $3 \times 3$ blocks (see (A.3)). In Appendix A we give the zero potential $K_B$ such that $H'_B = H_B + K_B$ is an $m$–potential from which one obtains the following results valid if $U$ is large enough.

Let

$$\mu_1 = -2\frac{t^2}{U} + (\frac{1}{2} - \cos\phi)\frac{t^4}{U^3}$$

$$\mu_2 = \begin{cases} -2\frac{t^2}{U} + (3 - \cos\phi)\frac{t^4}{U^3} & \text{if } \cos\phi \geq 0 \\ -2\frac{t^2}{U} + (3 + \frac{3}{2}\cos\phi)\frac{t^4}{U^3} & \text{if } \cos\phi \leq 0 \end{cases}$$

$$\mu_3 = \begin{cases} -2\frac{t^2}{U} + (3 + 3\cos\phi)\frac{t^4}{U^3} & \text{if } \cos\phi \geq 0 \\ \mu_2 & \text{if } \cos\phi \leq 0 \end{cases}$$

$$\mu_4 = -2\frac{t^2}{U} + (\frac{15}{2} + 3\cos\phi)\frac{t^4}{U^3} \qquad (3.11)$$

then

1. For $\mu < \mu_1$, $s_-$ is the unique ground state.

2. For $\mu_1 < \mu < \mu_2$, $H'_B$ is minimum for the configurations of fig. 2, and those obtained by rotations or reflections.

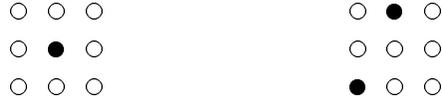

Figure 2: Configurations defining $\mathcal{S}_1$

3. For $\mu_2 < \mu < \mu_3$, $H'_B$ is minimum for the configurations of fig. 3.

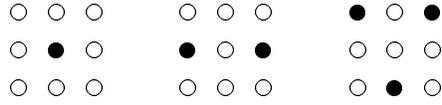

Figure 3: Configurations defining $\mathcal{S}_2$

4. For $\mu_3 < \mu < \mu_4$, $H'_B$ is minimum for the configurations of fig. 4.

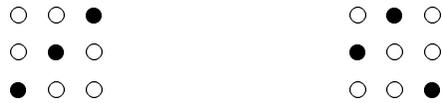

Figure 4: Configurations defining $\mathcal{S}_3$

5. For $\mu_4 < \mu \leq 0$, $H'_B$ is minimum for the configurations shown in fig. 5, and thus $\{s^1_{cb}, s^2_{cb}\}$ are the unique ground states.

From the above discussion we conclude that the only configurations which yield the minimum energy for each block $B$ are those represented on fig. 6, together with the empty and chessboard configurations, and those obtained by ion–hole inversion, and Proposition 3.1 is established.

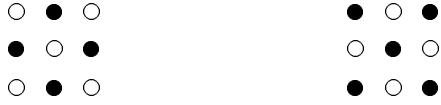

Figure 5: Configurations defining the chessboard configurations

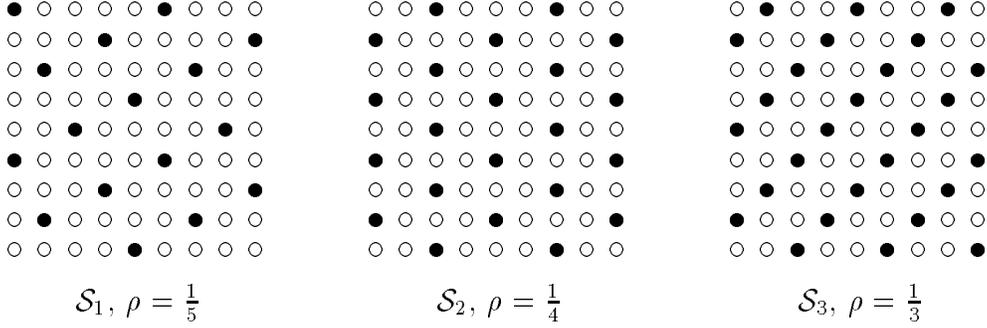

$\mathcal{S}_1$, $\rho = \frac{1}{5}$ $\qquad\qquad$ $\mathcal{S}_2$, $\rho = \frac{1}{4}$ $\qquad\qquad$ $\mathcal{S}_3$, $\rho = \frac{1}{3}$

Figure 6: Ground states appearing in the phase diagram for fermions on square lattice; the occupied sites are black, while the empty ones are white.

**Proposition 3.1** *For $U/t$ large enough, the ground states of the truncated hamiltonian at order 3 are the following $\mathcal{S}_j$ with density $\rho_j$ :*

$$\begin{aligned}
-\infty < \mu < \mu_1 & \quad \mathcal{S}_- = \{s_-\} & \rho_- = 0 \\
\mu_1 < \mu < \mu_2 & \quad \mathcal{S}_1 = \{s_1^\alpha\},\ \alpha = 1,\ldots,10 & \rho_1 = \frac{1}{5} \\
\mu_2 < \mu < \mu_3 & \quad \mathcal{S}_2 = \{s_2^\alpha\},\ \alpha = 1,\ldots,8 & \rho_2 = \frac{1}{4} \\
\mu_3 < \mu < \mu_4 & \quad \mathcal{S}_3 = \{s_3^\alpha\},\ \alpha = 1,\ldots,6 & \rho_3 = \frac{1}{3} \\
\mu_4 < \mu \leq 0 & \quad \mathcal{S}_{cb} = \{s_{cb}^1, s_{cb}^2\} & \rho_{cb} = \frac{1}{2}
\end{aligned}$$

*Note that $\mathcal{S}_2$ does not appear for $\phi \in ]\frac{\pi}{2}, \frac{3\pi}{2}[$.*

The phase diagram being symmetric with respect to $\mu = 0$, we have thus obtained the phase diagram for the truncated hamiltonian $H^{(3)}$ (fig. 7).

*b) Ground states for $H_{eff}$.*

We shall now show that the phase diagram for the effective hamiltonian is the same as the one for the truncated hamiltonian, except for domains of width of the order $t^6 U^{-5}$ along the separation lines. Let us first make the following observation: for a chemical potential $\mu$ in $]\mu_j, \mu_{j+1}[$ the energy of any configuration $s$ such that $s|_B \neq s_j^\alpha|_B$ will satisfy the inequality

$$H'_B(s) - H'_B(s_j^\alpha) \geq \tau > 0 \tag{3.12}$$

where $\tau$ is a function of $\mu$, $U$ and $\Phi$.

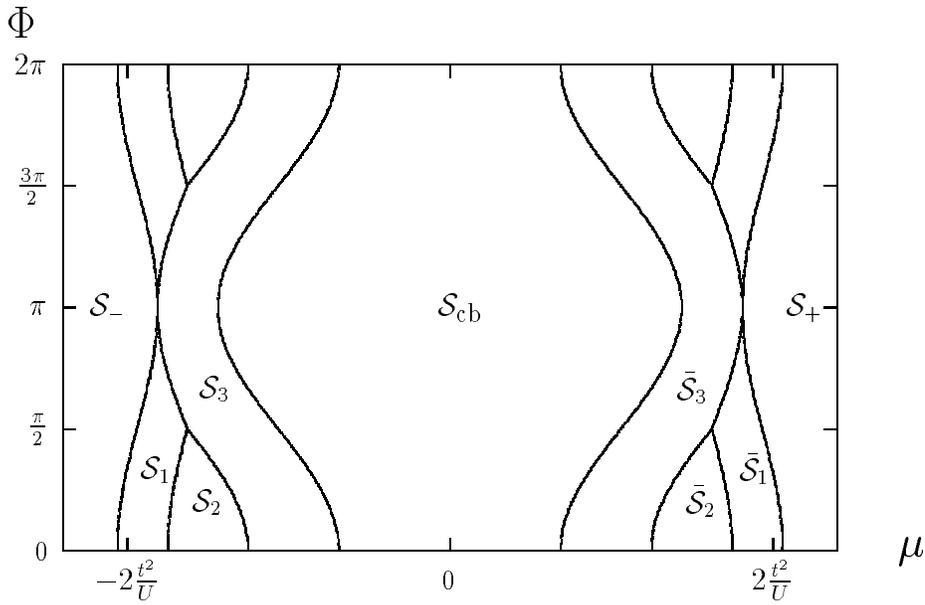

Figure 7: Fermions on a square lattice. Phase diagram for the hamiltonian truncated at order 3.

In the following we consider the case $\mu \in ]\mu_3, \mu_{cb}[$ for which the six ground state configurations are given by $\mathcal{S}_3$ and to simplify the notation we write $s^\alpha$ instead of $s_3^\alpha$. Let $s$ be a configuration and $Y$ a connected set of sites. We say that $Y$ is "*correct*" if there exists some $s^\alpha \in \mathcal{S}_3$ such that $s|_Y = s^\alpha|_Y$; moreover $Y$ is said "*$\alpha$-correct*" if there is a unique $s^\alpha$ satisfying this condition.

We then introduce the following notations :

$B_x$ is the $3 \times 3$ block centered in $x$,

$C_p(x)$ is the $p \times p$ block centered on $x$ if $p$ is odd, centered on $x + (\frac{1}{2}, \frac{1}{2})$ if $p$ is even. The integer $p$ is the lowest common multiple of $p_1, p_2$, the periods of $s$ in the directions 1 and 2. For example, $p = 3$ for $\mathcal{S}_3$. Note that if $C_p(x)$ is correct it is necessarily $\alpha$-correct with $\alpha = \alpha(x)$.

and for any configuration $s$

$\Gamma(s) = \{x \in \Lambda; B_x \text{ not-correct}\}$

$\mathcal{B}(s) = \{B_x \subset \Lambda; B_x \text{ not-correct}\} \qquad |\mathcal{B}(s)| = |\Gamma(s)|$

$\bar{\Gamma}(s) = \cup_{\substack{B \in \mathcal{B}(s) \\ C_p(x) \supset B}} C_p(x)$

It follows that $C_p(x)$ is not-correct if and only if $C_p(x) \subset \bar{\Gamma}(s)$. Therefore for any $y \notin \bar{\Gamma}(s)$, the condition $C_p(x) \ni y$ implies that $C_p(x)$ is $\alpha$-correct and all $C_p$ containing $y$ will be correct with the same $\alpha(x)$.

Let us decompose the complementary set $\bar{\Gamma}^c(s) = \Lambda \setminus \bar{\Gamma}(s)$ into maximal connected subsets

$$\bar{\Gamma}^c(s) = \cup_k I_k. \tag{3.13}$$

Then each $I_k$ is $\alpha$–correct, with $\alpha = \alpha(k)$.

With the definition of $\mathcal{B}(s)$, and (3.4), we have :

$$H_{eff}(s) - H_{eff}(s^\alpha) = \sum_{B \in \mathcal{B}(s)} (H'_B(s) - H'_B(s^\alpha))$$
$$+ \sum_{x \in \Lambda} \left( \bar{R}^{(3)}_x(s) - \bar{R}^{(3)}_x(s^\alpha) \right) \tag{3.14}$$

Using then the inequality (3.12) for the first sum and the expansion of $\bar{R}^{(3)}_x$ in terms of oriented paths, we obtain :

$$|H_{eff}(s) - H_{eff}(s^\alpha)| \geq \tau |\mathcal{B}(s)| - \frac{1}{p^2} \sum_{x \in \Lambda} \sum_{q \geq 5} \frac{t^{q+1}}{U^q}$$
$$\cdot \sum_{\substack{\omega \in \Omega_{q+1}(s) \\ \omega \cap C_p(x) \neq \phi}} |g_\omega(s) - g_\omega(s^{\alpha(x)})| \frac{|\text{supp } \omega \cap C_p(x)|}{|\text{supp } \omega|} \tag{3.15}$$

where because of (3.7) we have replaced $s^\alpha$ by $s^{\alpha(x)}$ (we set $\alpha(x) = \alpha$ if $C_p(x)$ is not correct).

All circuits $\omega$ such that $\omega \cap C_p(x) \neq \phi$, $\omega \cap \bar{\Gamma}(s) = \phi$, are $\alpha(x)$–correct, and thus

$$|H_{eff}(s) - H_{eff}(s^\alpha)| \geq \tau |\mathcal{B}(s)| - \frac{1}{p^2} \sum_{x \in \Lambda} \sum_{q \geq 5} \frac{t^{q+1}}{U^q}$$
$$\cdot \sum_{\substack{\omega \in \Omega_{q+1}(s) \\ \omega \cap C_p(x) \neq \phi, \omega \cap \bar{\Gamma}(s) \neq \phi}} |g_\omega(s) - g_\omega(s^{\alpha(x)})| \frac{|\text{supp } \omega \cap C_p(x)|}{|\text{supp } \omega|} \tag{3.16}$$

The function $g_\omega(s)$ is bounded by $\frac{1}{2N(\omega)}$. Then

$$|H_{eff}(s) - H_{eff}(s^\alpha)| \geq \tau |\mathcal{B}(s)| - \frac{1}{p^2} \sum_{q \geq 5} \frac{t^{q+1}}{U^q}$$
$$\cdot \sum_{\substack{\omega \in \Omega_{q+1}(s) \\ \omega \cap \bar{\Gamma}(s) \neq \phi}} \frac{1}{N(\omega)|\text{supp } \omega|} \left[ \sum_{x \in \Lambda} |\text{supp } \omega \cap C_p(x)| \right] \tag{3.17}$$

The last sum is bounded by $p^2 |\text{supp } \omega|$, thus

$$|H_{eff}(s) - H_{eff}(s^\alpha)| \geq \tau |\mathcal{B}(s)| - \sum_{q \geq 5} \frac{t^{q+1}}{U^q} \sum_{\substack{\omega = (x_1, \ldots, x_{q+1}) \\ x_1 \in \bar{\Gamma}(s)}} 1 \tag{3.18}$$

Let $K = |\cup_{C_p(x) \supset B} C_p(x)|$, ($K = 9$ in the case considered; $K = 25$ for $\mathcal{S}_2$, $K = 49$ for $\mathcal{S}_1$); then

$$|H_{eff}(s) - H_{eff}(s^\alpha)| \geq |\mathcal{B}(s)| \left( \tau - 9 \left( \frac{4t}{U} \right)^5 t(1 - \frac{4t}{U})^{-1} \right) \tag{3.19}$$

Therefore, for $\mu$ such that

$$\tau > 9 \left( \frac{4t}{U} \right)^5 t(1 - \frac{4t}{U})^{-1} \tag{3.20}$$

the set $\mathcal{S}_3$ is the set of ground states for the effective hamiltonian. In Appendix A we have shown the following result

**Proposition 3.2** *If $U/t > 387$ then there exist $\varepsilon_3, \delta_3$ such that for $\mu \in ]\mu_3 + \varepsilon_3, \mu_{cb} - \delta_3[$ the ground states of the effective hamiltonian are $\mathcal{S}_3$.*

For other intervals we proceed in a similar manner.

## 3.3 Bosons on the square lattice

From the expansion of the effective hamiltonian (2.28), one observes that at order 1 the results for hard–core bosons are the same as for fermions. At order 3 the truncated hamiltonian for hard–core bosons and fermions are identical if $\phi_P = \pi/2$, moreover $H^{(3)}_{\text{bosons}}(s; \phi) = H^{(3)}_{\text{fermions}}(s; \pi - \phi)$ if $s_P = 1$, for all $P$.

We can write (2.28) in the form

$$H^{(3)}(s) = \text{cte} + \sum_P H_P(s) + \frac{t^4}{8U^3} \sum_{|x-y|=2} s_x s_y. \tag{3.21}$$

Following the same method as for the fermions, one first shows that for $U$ large enough and $\mu \leq 0$, the ground states of $H^{(3)}$ cannot contain bonds $<x, y>$ with $s_x = s_y = 1$. Then one observes that, with $\mu = -\frac{t^2}{2U} + \delta \frac{t^4}{U^3}$

$$H_P \left( \begin{smallmatrix} \circ & \circ \\ \circ & \circ \end{smallmatrix} \right) = -\frac{t^2}{2U} + \frac{t^4}{8U^3}[4\delta - 3] \tag{3.22}$$

$$H_P \left( \begin{smallmatrix} \circ & \circ \\ \circ & \bullet \end{smallmatrix} \right) = -\frac{t^2}{2U} + \frac{t^4}{8U^3}[2\delta - 2\cos\phi] \tag{3.23}$$

$$H_P \left( \begin{smallmatrix} \bullet & \circ \\ \circ & \bullet \end{smallmatrix} \right) = -\frac{t^2}{2U} + \frac{t^4}{8U^3}[11 - 4\cos\phi] \tag{3.24}$$

Therefore for any configuration which does not contain bonds $<x, y>$ with $s_x = s_y = 1$, we have

$$\begin{aligned} H^{(3)}(s; \delta, \phi) &= H^{(3)}(s; \delta - \cos\phi, \phi = \frac{\pi}{2}) + \text{cte}(\phi) \\ &= H^{(3)}_{\text{fermions}}(s; \delta - \cos\phi, \phi = \frac{\pi}{2}) + \text{cte} \end{aligned} \tag{3.25}$$

(where cte($\phi$) is independent of $s$ and $\delta$) and thus the phase diagram for bosons with hard–core is directly obtained from the phase diagram of fermions with $\phi = \frac{\pi}{2}$, (fig. 8).

## 3.4 Fermions on the triangular lattice

As mentioned in the introduction the triangular lattice is especially interesting since it is not bipartite and the system does not have the usual particle–hole symmetry.

From the expansion (2.26) for the effective hamiltonian, one sees immediately that at the order $U^{-1}$ the truncated hamiltonian $H^{(1)}$ is the frustrated antiferromagnetic

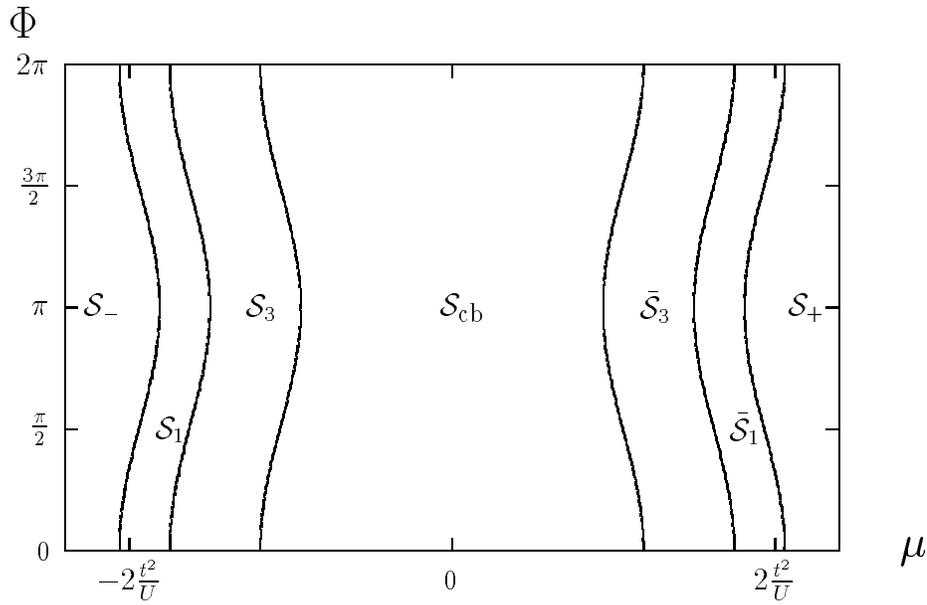

Figure 8: Hard-core bosons on a square lattice. Phase diagram of the hamiltonian truncated at order 3.

Ising model with magnetic field $-\frac{\mu}{2}$. Going to next order, which is $U^{-2}$ in this case, we will see that the frustration disappears.

At the order $U^{-2}$ we have :

$$H_{eff}(s) = H^{(2)}(s) + R^{(2)}(s) + \text{cte} \tag{3.26}$$
$$H^{(2)}(s) = \sum_{\Delta} H_{\Delta}(s)$$
$$H_{\Delta}(s) = -\left(\frac{\mu}{12} + \frac{t^3}{8U^2}\cos\phi_{\Delta}\right)\sum_{x \in \Delta} s_x + \frac{t^2}{8U}\sum_{<x,y> \subset \Delta} s_x s_y + \frac{3}{8}\frac{t^3}{U^2}\cos\phi_{\Delta}s_{\Delta}.$$

Computing $H_{\Delta}(s)$ for the eight configurations on the triangle $\Delta$, we have the following result; let us define

$$\mu_1 = -3\frac{t^2}{U} + 3\frac{t^3}{U^2}\cos\phi_{\Delta} \tag{3.27}$$

$$\mu_2 = -6\frac{t^3}{U^2}\cos\phi_{\Delta} \tag{3.28}$$

$$\mu_3 = 3\frac{t^2}{U} + 3\frac{t^3}{U^2}\cos\phi_{\Delta} \tag{3.29}$$

then the ground states of $H^{(2)}$ are given by (see fig. 9)

$$\mu < \mu_1 \quad \mathcal{T}_- = \{s_-\}, \quad \rho_- = 0$$
$$\mu_1 < \mu < \mu_2 \quad \mathcal{T}_5 = \{s_5^{\alpha}\}, \alpha = 1,2,3, \quad \rho_5 = \frac{1}{3}$$
$$\mu_2 < \mu < \mu_3 \quad \bar{\mathcal{T}}_5 = \{\bar{s}_5^{\alpha}\}, \alpha = 1,2,3, \quad \bar{\rho}_5 = \frac{2}{3}$$
$$\mu > \mu_3 \quad \mathcal{T}_+ = \{s_+\}, \quad \rho_+ = 1$$

where $s_5^{\alpha}$ is a configuration with one of the three sublattices occupied, and $\bar{s}_5^{\alpha}$ is a configuration with two sublattices occupied.

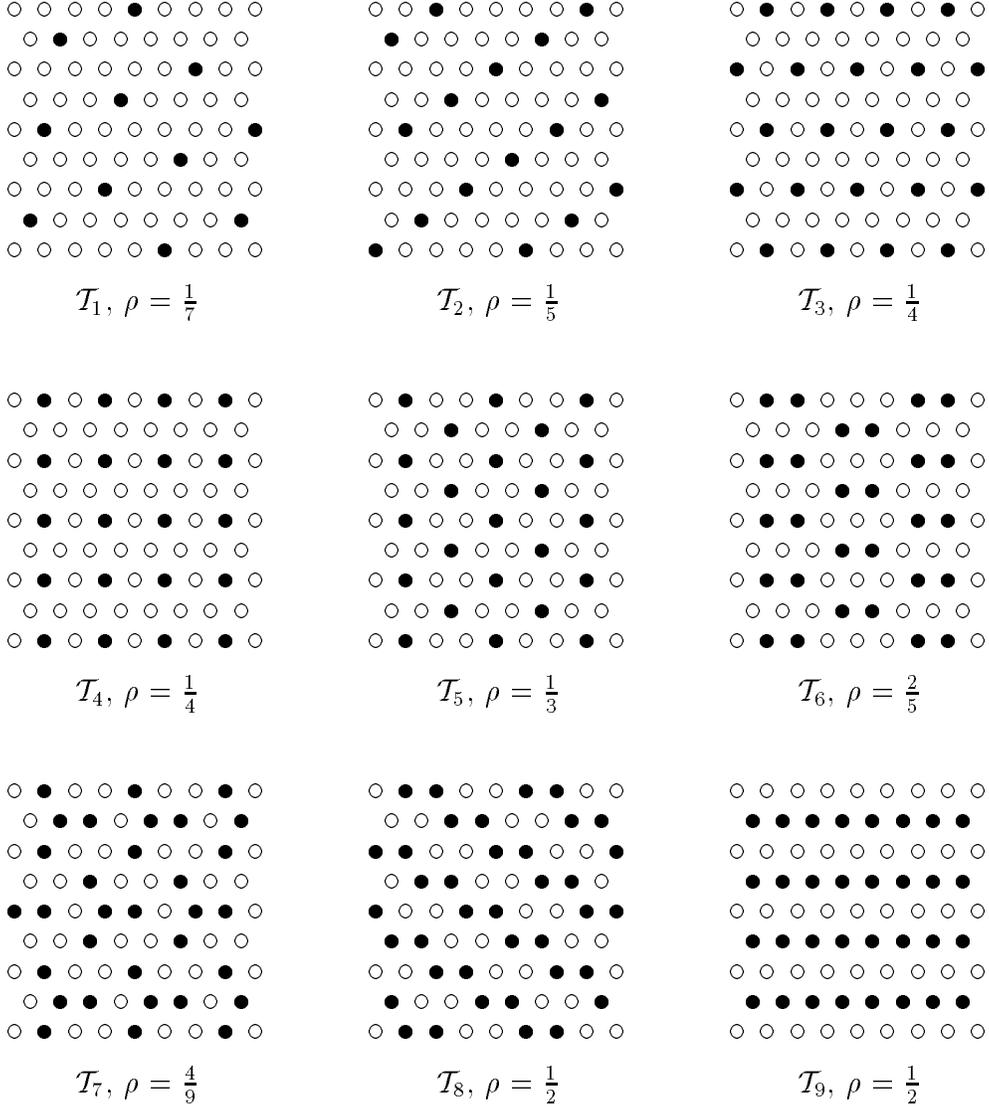

Figure 9: Ground states appearing in the phase diagram for fermions on triangular lattice; the occupied sites are black, while the empty ones are white.

The phase diagram for the truncated hamiltonian $H^{(2)}$ is represented on figure 10, which shows explicitely that there is no particle–hole symmetry. On the boundary the ground states are infinitely degenerate. Following the method described in section 3.2 one can then prove that for $U/t$ sufficiently large the phase diagram for the effective hamiltonian remains the same, except for domains with width of order $t^3/U^2$ around the boundaries.

Going to order 3 the problem of the ground states of the truncated hamiltonian $H^{(3)}$ becomes more difficult. In the Appendix we give zero–potentials from which one can deduce the ground states of $H^{(3)}$ for some domains of the $(\mu, \phi)$–plane, but we could not find a rigorous proof for all values $(\mu, \phi)$. Let us note however that except for the configurations $\mathcal{T}_2, \mathcal{T}_4$, all the configurations $\mathcal{T}_\alpha$, $\alpha = 1, \ldots, 9$ (fig. 9) do appear in the domains where the ground states have been rigourously established.

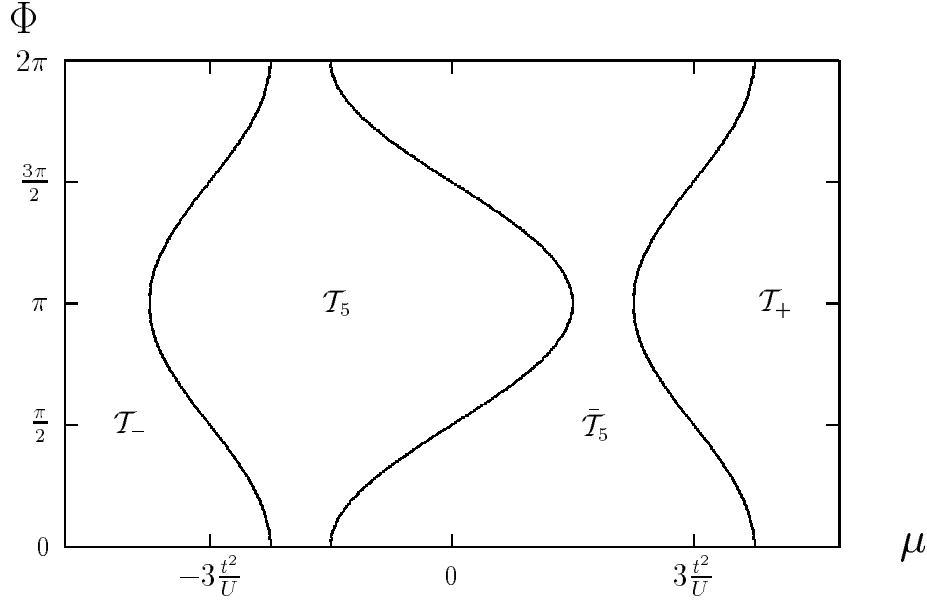

Figure 10: Fermions on a triangular lattice. Phase diagram of the hamiltonian truncated at order 2.

The results discussed in the Appendix are summarized as follows. Let us define the following functions of $\phi$

$$\mu_1 = -3\frac{t^2}{U} + 3\cos\phi\frac{t^3}{U^2} + (\frac{3}{4} - 3\cos 2\phi)\frac{t^4}{U^3}$$

$$\mu_2 = -3\frac{t^2}{U} + 3\cos\phi\frac{t^3}{U^2} + \begin{cases} (\frac{17}{4} - 3\cos 2\phi)\frac{t^4}{U^3} & \text{if } \cos 2\phi > 0 \\ (\frac{17}{4} + \frac{1}{2}\cos 2\phi)\frac{t^4}{U^3} & \text{if } \cos 2\phi < 0 \end{cases}$$

$$\mu_3 = -3\frac{t^2}{U} + 3\cos\phi\frac{t^3}{U^2} + (\frac{17}{4} - 2\cos 2\phi)\frac{t^4}{U^3} \text{ if } -\frac{1}{2} < \cos 2\phi < 0$$

$$\mu_4 = -3\frac{t^2}{U} + 3\cos\phi\frac{t^3}{U^2} + \begin{cases} (\frac{33}{4} + 9\cos 2\phi)\frac{t^4}{U^3} & \text{if } \cos 2\phi > 0 \\ (\frac{33}{4} + 6\cos 2\phi)\frac{t^4}{U^3} & \text{if } -\frac{1}{2} < \cos 2\phi < 0 \\ (\frac{27}{4} + 3\cos 2\phi)\frac{t^4}{U^3} & \text{if } \cos 2\phi < -\frac{1}{2} \end{cases}$$

$$\mu_5 = -6\cos\phi\frac{t^3}{U^2} + \begin{cases} 0 & \text{if } \cos 2\phi > \frac{3}{4} \\ (-\frac{27}{8} + \frac{9}{2}\cos 2\phi)\frac{t^4}{U^3} & \text{if } \frac{5}{12} < \cos 2\phi < \frac{3}{4} \\ (-\frac{21}{4} + 9\cos 2\phi)\frac{t^4}{U^63} & \text{if } \cos 2\phi < \frac{5}{12} \end{cases}$$

$$\mu_6 = -6\cos\phi\frac{t^3}{U^2} + (-\frac{9}{16} - \frac{9}{4}\cos 2\phi)\frac{t^4}{U^3} \text{ if } \frac{1}{4} < \cos 2\phi < \frac{5}{12}$$

$$\mu_7 = -6\cos\phi\frac{t^3}{U^2} + \begin{cases} 0 & \text{if } \cos 2\phi > \frac{1}{2} \\ (-\frac{9}{4} + \frac{9}{2}\cos 2\phi)\frac{t^4}{U^3} & \text{if } \frac{1}{4} < \cos 2\phi < \frac{1}{2} \\ (-\frac{3}{2} + \frac{3}{2}\cos 2\phi)\frac{t^4}{U^3} & \text{if } 0 < \cos 2\phi < \frac{1}{4} \\ (-\frac{3}{2} + 9\cos 2\phi)\frac{t^4}{U^3} & \text{if } \cos 2\phi < 0 \end{cases}$$

(3.30)

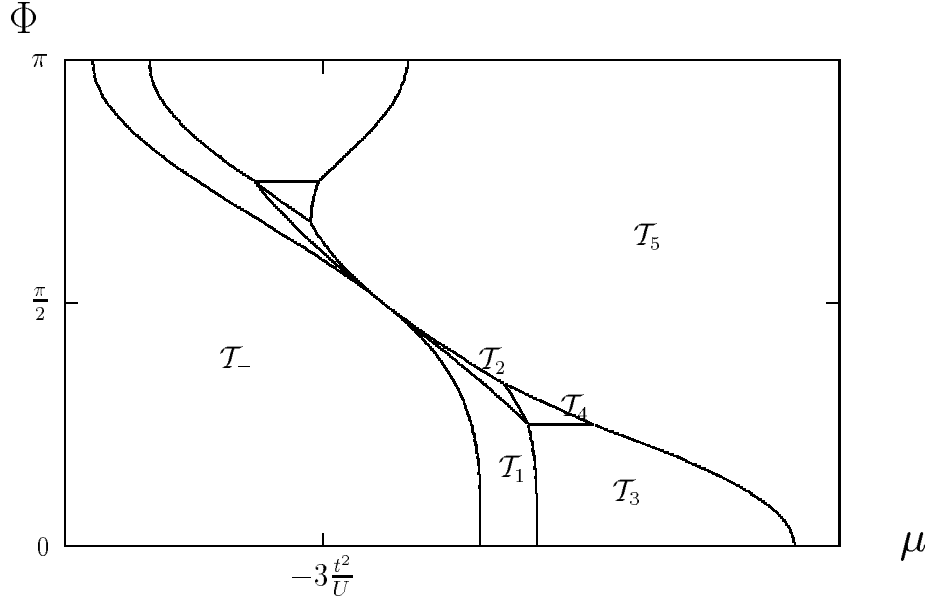

Figure 11: Fermions on a triangular lattice. Phase diagram for the hamiltonian truncated at order 3, for chemical potentials corresponding to low densities.

With the configurations $\mathcal{T}_\alpha, \alpha = 1, \ldots, 9$ represented on fig. 9, the ground states of $H^{(3)}$ are given around $\mu = -3\frac{t^2}{U}$ by

$\mathcal{T}_-,\ \rho_- = 0 \quad \mu < \mu_1$

$\mathcal{T}_1,\ \rho_1 = \frac{1}{7} \quad \mu_1 < \mu < \mu_2$

$\mathcal{T}_2,\ \rho_2 = \frac{1}{5} \quad \mu_2 < \mu < \begin{cases} \mu_3 & \text{if } -\frac{1}{2} < \cos 2\phi < 0 \\ \mu_4 & \text{if } \cos 2\phi < -\frac{1}{2} \end{cases}$

$\mathcal{T}_3,\ \rho_3 = \frac{1}{4} \quad \mu_2 < \mu < \mu_4 \text{ and } \cos 2\phi > 0$

$\mathcal{T}_4,\ \rho_4 = \frac{1}{4} \quad \mu_3 < \mu < \mu_4 \text{ and } -\frac{1}{2} < \cos 2\phi < 0$

$\mathcal{T}_5,\ \rho_5 = \frac{1}{3} \quad \mu_4 < \mu < \mu_5$

while for values of the chemical potential around 0,

$\mathcal{T}_6,\ \rho_6 = \frac{2}{5} \quad \mu_5 < \mu < \begin{cases} \mu_6 & \text{if } \frac{1}{4} < \cos 2\phi < \frac{1}{2} \\ \mu_7 & \text{if } \cos 2\phi < \frac{1}{4} \end{cases}$

$\mathcal{T}_7,\ \rho_7 = \frac{4}{9} \quad \begin{cases} \mu_5 < \mu < \mu_7 & \text{if } \cos 2\phi > \frac{5}{12} \\ \mu_6 < \mu < \mu_7 & \text{if } \frac{1}{4} < \cos 2\phi < \frac{5}{12} \end{cases}$

$\mathcal{T}_8,\ \rho_8 = \frac{1}{2} \quad \mu_7 < \mu < \bar{\mu}_7 \text{ and } \cos 2\phi > 0$

$\mathcal{T}_9,\ \rho_9 = \frac{1}{2} \quad \mu_7 < \mu < \bar{\mu}_7 \text{ and } \cos 2\phi < 0$

where $\bar{\mu}_j$ is defined from $\mu_j$ as follows : the coefficient of $\frac{t^3}{U^2}$ is identical, while the ones of $\frac{t^2}{U}$ and $\frac{t^4}{U^3}$ are opposite.

The phase diagrams are represented on figures 10, 11 and 12.

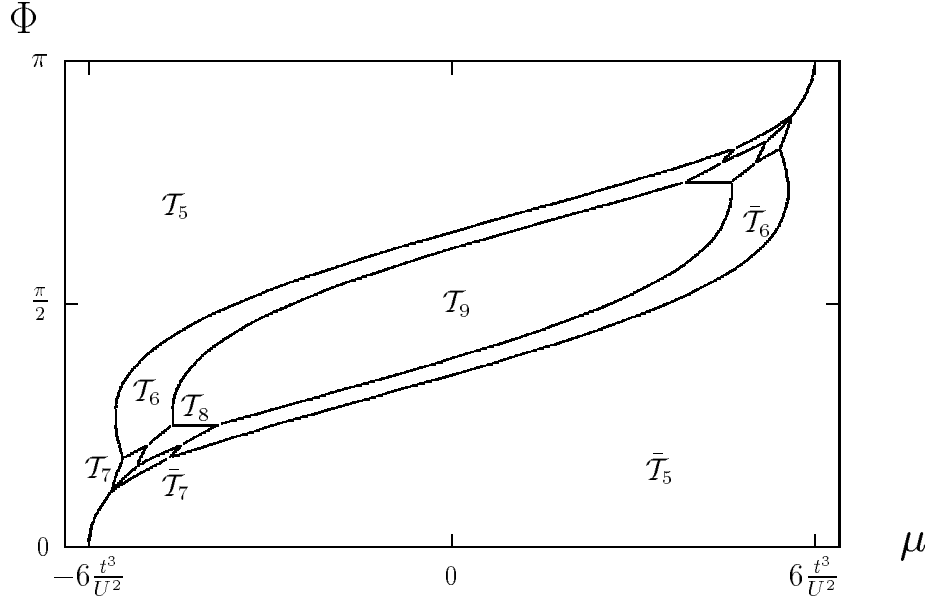

Figure 12: Fermions on a triangular lattice. Phase diagram for the hamiltonian truncated at order 3, for chemical potentials corresponding to densities close to one half.

## 3.5 Bosons on triangular lattice

From the expansion (2.29) one observes that at the order $U^{-1}$, the truncated hamiltonian for hard–core bosons is the same as for fermions (antiferromagnetic Ising model with field). Going to the next order, i.e. $U^{-2}$, we have :

$$H^{(2)}(s) = \sum_{\Delta} \left\{ -\frac{\mu}{12} \sum_{x \in \Delta} s_x + \left( \frac{t^2}{8U} + \frac{t^3}{8U^2} \cos \phi \right) \sum_{<x,y> \subset \Delta} s_x s_y \right\} \qquad (3.31)$$

from which one concludes immediately that the ground states of $H^{(2)}$ are the following:

$$\mu < -3\frac{t^2}{U} - 3 \cos \phi \frac{t^3}{U^2} \qquad \mathcal{T}_- = \{s_-\}, \qquad \rho_- = 0$$

$$-3\frac{t^2}{U} - 3 \cos \phi \frac{t^3}{U^2} < \mu < 0 \quad \mathcal{T}_5 = \{s_5^\alpha\} \; \alpha = 1,2,3, \quad \rho_5 = \frac{1}{3}$$

$$0 < \mu < 3\frac{t^2}{U} + 3 \cos \phi \frac{t^3}{U^2} \quad \bar{\mathcal{T}}_5 = \{\bar{s}_5^\alpha\} \; \alpha = 1,2,3, \quad \bar{\rho}_5 = \frac{2}{3}$$

$$\mu > 3\frac{t^2}{U} + 3 \cos \phi \frac{t^3}{U^2} \qquad \mathcal{T}_+ = \{s_+\}, \qquad \rho_+ = 1$$

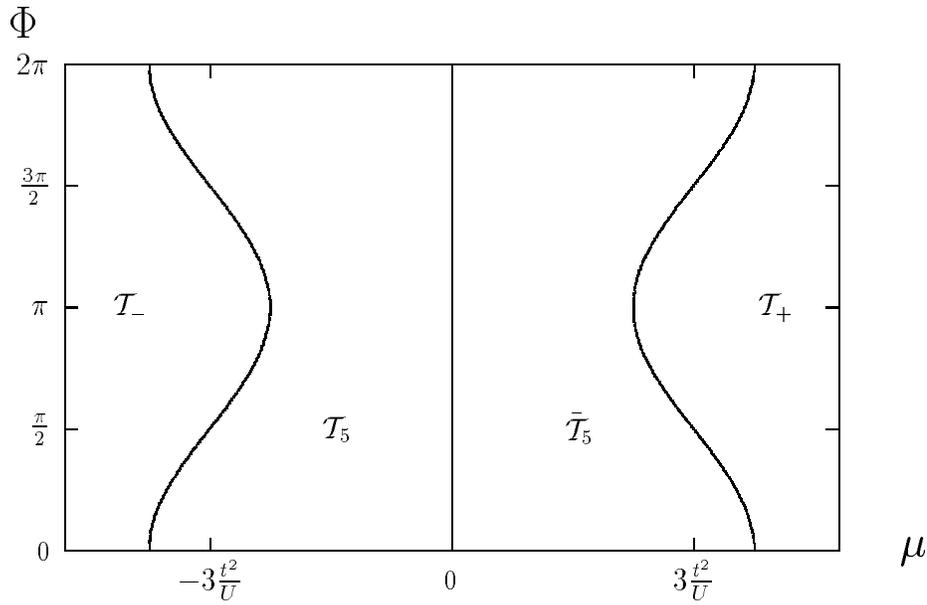

Figure 13: Hard–core bosons on a triangular lattice. Phase diagram for the hamiltonian truncated at order 2.

Then except for domains of order $\frac{t^4}{U^3}$ the phase diagram of the effective hamiltonian is the one presented on fig. 13.

In the Appendix we then show that it is possible to find zero–potentials such that the truncated hamiltonian at order 3 is an $m$–potential. We then obtain the following result. Let

$$\mu_1 = -3\frac{t^2}{U} - 3\cos\phi\frac{t^3}{U^2} + (\frac{3}{4} - 3\cos 2\phi)\frac{t^4}{U^3} \tag{3.32}$$

$$\mu_2 = -3\frac{t^2}{U} - 3\cos\phi\frac{t^3}{U^2} + (\frac{17}{4} - 3\cos 2\phi)\frac{t^4}{U^3} \tag{3.33}$$

$$\mu_3 = -3\frac{t^2}{U} - 3\cos\phi\frac{t^3}{U^2} + (\frac{33}{4} - 3\cos 2\phi)\frac{t^4}{U^3} \tag{3.34}$$

$$\mu_4 = (-\frac{9}{2} - 3\cos 2\phi)\frac{t^4}{U^3} \tag{3.35}$$

$$\mu_5 = \begin{cases} (-2 - 3\cos 2\phi)\frac{t^4}{U^3} & \text{if } \cos 2\phi \geq 0 \\ (-2 - \frac{1}{2}\cos 2\phi)\frac{t^4}{U^3} & \text{if } \cos 2\phi \leq 0 \end{cases} \tag{3.36}$$

Then the description of the phase diagram of the truncated hamiltonian at order 3 is as follows

$$\begin{aligned}
\mathcal{T}_-, & \quad \rho_- = 0 & \mu < \mu_1 \\
\mathcal{T}_1, & \quad \rho_1 = \tfrac{1}{7} & \mu_1 < \mu < \mu_2 \\
\mathcal{T}_{1/4}^*, & \quad \rho = \tfrac{1}{4} & \mu_2 < \mu < \mu_3 \\
\mathcal{T}_5, & \quad \rho_5 = \tfrac{1}{3} & \mu_3 < \mu < \mu_4 \\
\mathcal{T}_6, & \quad \rho_6 = \tfrac{2}{5} & \mu_4 < \mu < \mu_5
\end{aligned}$$

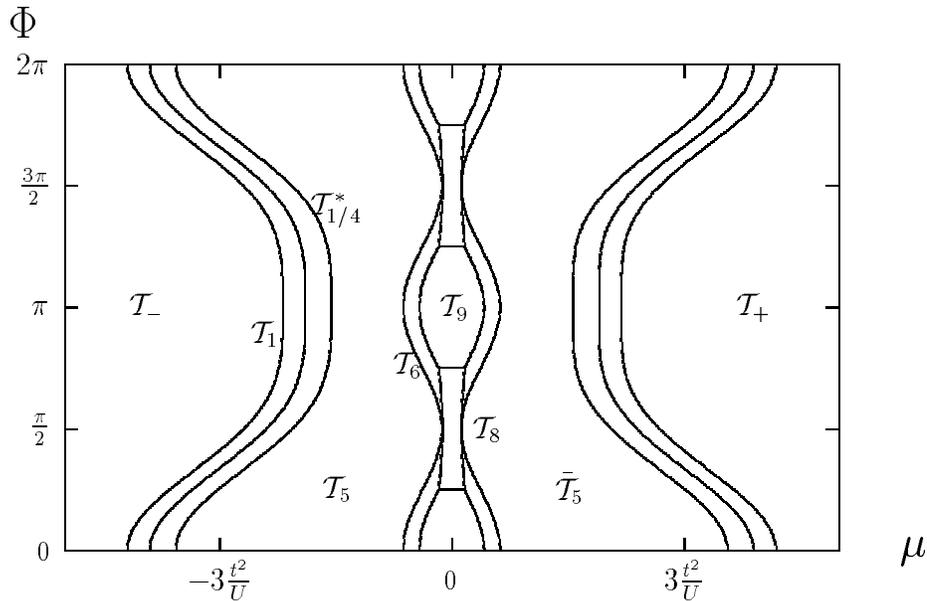

Figure 14: Hard–core bosons on a triangular lattice. Phase diagram of the hamiltonian truncated at order 3.

$\mathcal{T}_8$, $\rho_8 = \frac{1}{2}$ $\mu_5 < \mu \leq 0$ and $\cos 2\phi < 0$
$\mathcal{T}_9$, $\rho_9 = \frac{1}{2}$ $\mu_5 < \mu \leq 0$ and $\cos 2\phi > 0$

where $\mathcal{T}_{1/4}^*$ is the set of all configurations with density $\frac{1}{4}$ and the constraint that with respect to a lattice direction : each two lines is empty, while the others have alternatively one ion and one empty site ($\log|\mathcal{T}_{1/4}^*| = O(|\Lambda|^{\frac{1}{2}})$). This set is probably not a set of ground states for the effective hamiltonian.

The phase diagram is symmetric around $\mu = 0$; therefore the separation lines for $\mu \geq 0$ are derived immediately from $\mu_1, \ldots, \mu_5$.

## 4 Flux phase problem

### 4.1 Minimum flux for a given configuration of ions

In this section we consider the following problem : given a configuration $s$ of ions, what are the fluxes $\Phi^{\min}(s)$ across the elementary cells $C$ which minimize the effective hamiltonian ? As in the previous sections, $\Lambda$ is either the square or the triangular lattice, and the cells $C$ are either plaquettes $P$ or triangles $\Delta$. The proofs are based on the $U^{-1}$ expansion, thus the chemical potential for the electrons must lie in the gap and $U$ has to be sufficiently large. We take again $\mu_e = 0$ and write $\mu_i = \mu$.

The case of bosons was treated in section 2.4. From now on we consider only fermions systems.

**Theorem 4.1**   1. For the configurations $s = s_+$ or $s_-$, the effective hamiltonian is independent of the fluxes.

2. Let $s \notin \{s_+, s_-\}$ satisfying the condition that for any site $x$ there exists a site $y$ with $|x - y| \leq d(s) < \infty$ and $s_x = -s_y$. Then there exists a positive number $U_0$ (depending on $s$) such that for $U \geq U_0$, the minimizing fluxes are given by $\phi_C^{\min}(s) = \bar{\phi}_C(s)$ where

$$\text{square lattice}: \quad \bar{\phi}_P(s) = \begin{cases} \pi & \text{if 2 sites of } P \text{ are occupied} \\ 0 & \text{otherwise} \end{cases} \quad (4.1)$$

$$\text{triangular lattice}: \quad \bar{\phi}_\Delta = \begin{cases} \pi & \text{if 0 or 1 site of } \Delta \text{ is occupied} \\ 0 & \text{if 2 or 3 sites of } \Delta \text{ are occupied} \end{cases} \quad (4.2)$$

The detailed proof of theorem 4.1 can be found in Appendix B. Here let us outline the main ideas.

1. If $s = s_+$ or $s_-$, $\Omega(s)$ is empty and the effective hamiltonian does not depend on the fluxes. This is not surprising, because we took the chemical potential of the electrons to be in the gap, and therefore the number of electrons is either 0 or $|\Lambda|$; in the first case there is no electron to feel the magnetic field, and in the second case the electrons cannot hop because there is no free lattice sites; then the phases of the hopping coefficients have no influence.

2. Given $s \notin \{s_+, s_-\}$ defined on the square lattice, let $\mathcal{P}$ denote the set of plaquettes which are neither completely occupied nor empty. From (2.17) the contribution involving fluxes in the truncated hamiltonian $H^{(3)}(s)$ is

$$\frac{2}{8}\frac{t^4}{U^3} \sum_{P \in \mathcal{P}} (-1)^m \cos\phi_P \frac{2}{(m-1)!(3-m)!} = \frac{2}{8}\frac{t^4}{U^3} \sum_{P \in \mathcal{P}} \cos\varphi_P \frac{2}{(m-1)!(3-m)!} \quad (4.3)$$

where

$$\varphi_P = \phi_P + \frac{1}{2}(s_P - 1)\pi; \qquad (-1)^m = s_P \quad (4.4)$$

Therefore the fluxes which minimize $H^{(3)}(s)$ are $\phi_P = \frac{1}{2}(s_P + 1)\pi$ if $P \in \mathcal{P}$ and $\phi_P$ arbitrary if $P \notin \mathcal{P}$.

We then consider the truncated hamiltonian $H^{(5)}(s)$. Let $\mathcal{R}$ denote the set of rectangles (two plaquettes) which are neither completely filled nor empty. We decompose $\mathcal{R} = \mathcal{R}_1 \cup \mathcal{R}_2$ with $\mathcal{R}_1 = \{R_1 = P_0 \cup P'; P_0 \notin \mathcal{P}, P' \in \mathcal{P}\}$ and $\mathcal{R}_2 = \{R_2 = P' \cup P''; P' \in \mathcal{P}, P'' \in \mathcal{P}\}$. The contribution involving fluxes in $H^{(5)}(s)$ is

$$\frac{2}{8}\frac{t^4}{U^3} \left[ \sum_{R_1 \in \mathcal{R}_1} \left\{ \frac{1}{4}\cos\varphi_{P'} \left[ \frac{2}{(m'-1)!(3-m')!} + \frac{t^2}{U^2}c_1 \right] \right. \right.$$
$$\left. - \frac{t^2}{U^2}c_2 \cos\varphi_{P_0} + \frac{t^2}{4U^2}\cos(\varphi_{P'} + \varphi_{P_0}) \right\}$$
$$+ \sum_{R_2 \in \mathcal{R}_2} \left\{ \frac{1}{4}\cos\varphi_{P'} \left[ \frac{2}{(m'-1)!(3-m')!} + \frac{t^2}{U^2}c_1' \right] \right.$$
$$\left. \left. + \frac{1}{4}\cos\varphi_{P''} \left[ \frac{2}{(m'-1)!(3-m')!} + \frac{t^2}{U^2}c_1'' \right] + \frac{t^2}{4U^2}s_{xy}\cos(\varphi_{P'} + \varphi_{P''}) \right\} \right] \quad (4.5)$$

where $xy = P' \cap P''$ and we used the fact that $s_{P'}s_{P''}\cos(\phi_{P'} + \phi_{P''}) = \cos(\varphi_{P'} + \varphi_{P''})$. We note also that $|c_1|$, $|c_1'|$, $|c_1''|$ are smaller than 3 and $C - 2 \geq 0$.

Therefore for $U > 4t$ the first term is minimum for $\varphi_{P'} = \frac{1}{2}(s_{P'} + 1)\pi$ as before and $\varphi_{P_0} = 0$. On the other hand the second term is of the form

$$\alpha \cos \varphi' + \beta \cos \varphi'' + \varepsilon \frac{t^2}{U^2} \cos(\varphi' + \varphi'')$$

where $\alpha, \beta \geq \frac{1}{4}(1 - 3\frac{t^2}{U^2})$ and $|\varepsilon| = 1$. Therefore for $U > 4t$ this term is minimum for $\varphi_{P'} = \varphi_{P''} = \pi$, as was obtained already at the previous order. Finally the fluxes on those plaquettes $P$ such that $P \cap \mathcal{R} = \emptyset$ remain arbitrary. We then proceed by induction to determine the value of $\phi_P$ for the remaining plaquettes.

## 4.2 Flux phase problem

We first discuss the flux phase problem for the truncated hamiltonian at order 3. For a given configuration $s$, the fluxes $\{\tilde{\phi}_C(s)\}$ which minimize $H^{(3)}(s, \mu, \{\phi_C\})$ are given by (4.1) and (4.2) when $C$ is not filled or empty, and arbitrary otherwise.

Introducing appropriate zero–potentials (see Appendix C) we may write $H^{(3)}(s; \mu; \{\tilde{\phi}_C(s)\})$ as an $m$–potential

$$H^{(3)}(s; \mu; \{\tilde{\phi}_C(s)\}) = \sum_B H'_B(s; \mu; \{\tilde{\phi}_C(s)\}) \tag{4.6}$$

and one can then find its phase diagram, and thus the flux phase problem is solved for the truncated hamiltonian at order 3.

**Theorem 4.2** *For $U$ sufficiently large, the configurations of ions and the set of fluxes which minimize the truncated hamiltonian at order 3 are the following (fig. 15 and 16).*

*i) for the square lattice*

$$\mu < \mu_1 = -2\frac{t^2}{U} - \frac{1}{2}\frac{t^4}{U^3} \quad \mathcal{S}_-, \quad \rho_- = 0, \quad \tilde{\phi}_P \text{ is arbitrary}$$

$$\mu_1 < \mu < \mu_3 = -2\frac{t^2}{U} + 2\frac{t^4}{U^3} \quad \mathcal{S}_1, \quad \rho_1 = \frac{1}{5}, \quad \tilde{\phi}_P \text{ is 0 or arbitrary}$$

$$\mu_3 < \mu < \mu_{cb} = -2\frac{t^2}{U} + \frac{13}{2}\frac{t^4}{U^3} \quad \mathcal{S}_3, \quad \rho_3 = \frac{1}{3}, \quad \tilde{\phi}_P \text{ is 0 or } \pi$$

$$\mu_{cb} < \mu \leq 0 \quad \mathcal{S}_{cb}, \quad \rho_{cb} = \frac{1}{2}, \quad \tilde{\phi}_P \text{ is } \pi$$

*Using the particle–hole symmetry the phase diagram for $\mu \geq 0$ is obtained by replacing $s_\alpha$ by $-s_\alpha$, without changing the fluxes;*

*ii) for the triangular lattice,*

$$\begin{aligned}
\mu < \mu_1 &= -3\frac{t^2}{U} - 3\frac{t^3}{U^2} - \frac{9}{4}\frac{t^4}{U^3} & \mathcal{T}_-, \quad \rho_- = 0, \quad \tilde{\phi}_\Delta \text{ is arbitrary} \\
\mu_1 < \mu < \mu_2 &= -3\frac{t^2}{U} - 3\frac{t^3}{U^2} + \frac{5}{4}\frac{t^4}{U^3} & \mathcal{T}_1, \quad \rho_1 = \frac{1}{7}, \quad \tilde{\phi}_\Delta \text{ is } \pi \text{ or arbitrary} \\
\mu_2 < \mu < \mu_3 &= -3\frac{t^2}{U} - 3\frac{t^3}{U^2} + \frac{69}{4}\frac{t^4}{U^3} & \mathcal{T}_3, \quad \rho_3 = \frac{1}{4}, \quad \tilde{\phi}_\Delta \text{ is } \pi \\
\mu_3 < \mu < \mu_5 &= -\frac{15}{2}\frac{t^4}{U^3} & \mathcal{T}_5, \quad \rho_5 = \frac{1}{3}, \quad \tilde{\phi}_\Delta \text{ is } \pi \\
\mu_5 < \mu < \mu_6 &= -5\frac{t^4}{U^3} & \mathcal{T}_6, \quad \rho_6 = \frac{2}{5}, \quad \tilde{\phi}_\Delta \text{ is } 0 \text{ or } \pi \\
\mu_6 < \mu &\leq 0 & \mathcal{T}_9, \quad \rho_9 = \frac{1}{2}, \quad \tilde{\phi}_\Delta \text{ is } 0 \text{ or } \pi
\end{aligned}$$
(4.7)

*The phase diagram for $\mu \geq 0$ is obtained by replacing $\mu_\alpha$, $s_\alpha$, $\phi_\alpha$ by $-\mu_\alpha$, $-s_\alpha$, $\phi_\alpha - \pi$.*

*Remark :* for the square lattice, the configurations $\mathcal{S}_2$ do not appear except at the value $\mu = \mu_3$ which is infinitely degenerate. In the case of the triangular lattice, the configurations $\mathcal{T}_2$, $\mathcal{T}_4$, $\mathcal{T}_7$ and $\mathcal{T}_8$ do not appear.

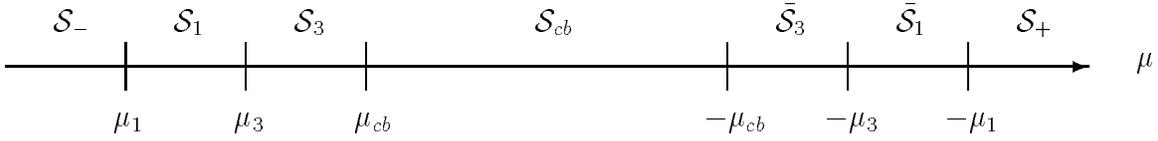

Figure 15: Flux phase problem on square lattice. Phase diagram of the hamiltonian truncated at order 3.

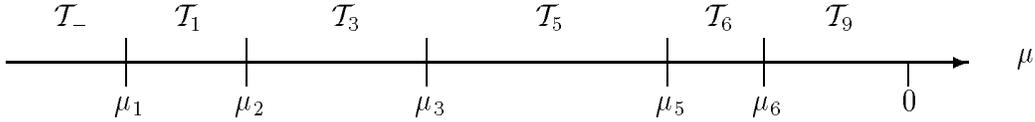

Figure 16: Phase diagram of the flux phase problem on triangular lattice.

Following the ideas developped for the case of constant fluxes, it is possible to define subintervals $]\mu'_j, \mu'_{j+1}[$ where a suitable Peierls condition can be obtained for $H^{(3)}$, and to prove that the phase diagram for the effective hamiltonian is the same as the phase diagram for $H^{(3)}$ up to domains of width of the order $\frac{t^5}{U^4}$ (or $\frac{t^6}{U^5}$ in the case of the square lattice) around the boundary points.

# 5  Conclusion

The phase diagram at finite temperature has been obtained in [12, 16], only for the square lattice and around the symmetry point. There the two chessboard ground states are related by a particle–hole or translational symmetry and therefore a Peierls argument can be applied. For other ground states this is not the case and Pirogov–Sinai theory has to be used[†]. Recent developments of this theory allowing general long–range (and also quantum) interactions may be found in [4] and [3]. Using these results it is expected that the phase diagram is stable at low temperatures.

The present work is limited to the large $U$ limit. For small $U$ the phase diagram changes drastically. This is the case in one and two dimensions where a very rich structure has been uncovered [9, 5, 27]. We expect that values different from 0 or $\pi$ will appear for the optimal flux as $U$ is lowered.

In the case of hard–core bosons when $U = 0$, we get the quantum $XY$ model. This model is known to display off–diagonal long–range order (ODLRO) at half–filling in the ground state [13]. On the other hand as soon as $U > 0$ the ground state configuration for the nuclei is a chessboard (on a square lattice) [22]. For large enough $U$ it seems clear that the bosons will form a charge density wave of period $\sqrt{2}$. An interesting question is whether the critical value $U_c$ for the disappearance of ODLRO is $U_c = 0$ or $U_c > 0$.

Finally we wish to point out the link between the Falicov–Kimball model and Hartree–Fock theory for the Hubbard model. If one minimizes the ground state (or free energy) for the repulsive Hubbard model, only over the set of Hartree–Fock states, then one finds that the mathematical structure of this variational problem is very similar to the one for the Hubbard model. This fact has been exploited in great detail recently [2]. The results obtained here on the ground states for rational filling factors of the form $\frac{p}{q}$ may shed some light on the Hartree–Fock states that one should look for in the Hubbard model.

We hope to come back to these questions in the future.

## Acknowledgements

We would like to thank J. L. Lebowitz for encouragement and useful suggestions concerning the triangular lattice, and V. Bach, J. Miękisz and S. Miracle–Solé for stimulating discussions. It is a pleasure to thank M. Merkli who participated at the initial stages of this work. This work was partly supported by the Swiss national foundation for science under grant 20-40'650.94.

---

[†]The reader interested in a pedagogical introduction to these ideas may look in [14]

# A $m$–potentials

## A.1 Square lattice

Following the method of [8] using plaquettes and bonds $<xyz>$, it is straightforward to show that for $U > 4t$, one has the following property : for $\mu < \mu_1$, $s_-$ is the only ground state of $H^{(3)}$, and for $0 \geq \mu \geq \mu_{cb}$, $\mathcal{S}_{cb} = \{s^1_{cb}, s^2_{cb}\}$ are the only ground states of $H^{(3)}$, where

$$\mu_1 = -2\frac{t^2}{U} + \frac{t^4}{U^3}(\frac{1}{2} - \cos\phi)$$

$$\mu_{cb} = -2\frac{t^2}{U} + \frac{t^4}{U^3}(\frac{15}{2} + 3\cos\phi)$$

To obtain the ground states of $H^{(3)}$ in the remaining interval $\mu \in [\mu_1, \mu_{cb}]$, we first note that for $U > 4t$ and $\mu < \mu_{cb}$ the ground state configurations cannot contain bonds $<x,y>$ with $s_x = s_y = 1$.

Introducing the parameter $\delta$ defined by

$$\mu = -2\frac{t^2}{U} + \frac{t^4}{U^3}\delta, \tag{A.1}$$

we write the truncated hamiltonian in the form

$$H^{(3)}(s) = \frac{t^2}{U} \sum_{\substack{<x,y> \\ s_x=s_y=+1}} 1 + \frac{t^4}{2U^3} \sum_{B \subset \Lambda} H_B(s;\mu) + \text{cte} \tag{A.2}$$

where $B$ are $3 \times 3$ blocks, centered on $x(B)$, and

$$H_B(s;\mu) = -\delta(s_{x(B)} + 1) - \frac{1}{48}(7 + 2\cos\phi)\sum_{<x,y>\subset B} s_x s_y$$

$$+ \frac{1}{32}(4 - \cos\phi)\sum_{\substack{(x,y)\subset B \\ |x-y|=\sqrt{2}}} s_x s_y + \frac{1}{12}\sum_{\substack{(x,y)\subset B \\ |x-y|=2}} s_x s_y + \frac{1}{32}\cos\phi\sum_{P\subset B}(1 + 5s_P) \tag{A.3}$$

As explained in the introduction we then introduce zero potentials $K_B$ such that $H'_B = H_B + K_B$ is minimum if and only if the configurations on $B$ are either $\mathcal{S}_1, \mathcal{S}_2$ or $\mathcal{S}_3$ depending on the value of $\mu$ (i.e. $\delta$). Numbering the nine sites in a block $B$ from left to right, starting down and ending up, let

$$K_B = \sum_{i=1}^{4} \alpha_i k_B^{(i)} \tag{A.4}$$

where the $\alpha_i$ are given in Table 1 and

$$k_B^{(1)} = -s_2 - s_4 - s_6 - s_8 + 4s_5 \tag{A.5}$$
$$k_B^{(2)} = -s_1 - s_3 - s_7 - s_9 + 4s_5 \tag{A.6}$$
$$k_B^{(3)} = -s_1 s_3 + 2s_4 s_6 - s_7 s_9 - s_1 s_7 + 2s_2 s_8 - s_3 s_9 \tag{A.7}$$
$$k_B^{(4)} = -s_2 s_4 - s_2 s_6 - s_4 s_8 - s_6 s_8 + s_1 s_5 + s_3 s_5 + s_5 s_7 + s_5 s_9 \tag{A.8}$$

These potentials $k_B^{(i)}$ are invariant by reflection and rotation.

Doing explicit calculations, we find that the only configurations which minimize $H_B'$ are the following :

If $\cos \phi \geq 0$, then for

$\frac{1}{2} - \cos \phi < \delta < 3 - \cos \phi$, it is $\{s_1^\alpha\}$

$3 - \cos \phi < \delta < 3 + 3\cos \phi$, it is $\{s_2^\alpha\}$

$3 + 3\cos \phi < \delta < \frac{15}{2} + 3\cos \phi$, it is $\{s_3^\alpha\}$

If $\cos \phi \leq 0$, then for

$\frac{1}{2} - \cos \phi < \delta < 3 + \frac{3}{2}\cos \phi$, it is $\{s_1^\alpha\}$

$3 + \frac{3}{2}\cos \phi < \delta < \frac{15}{2} + 3\cos \phi$, it is $\{s_3^\alpha\}$

which establishes the Proposition 3.1.

We then have to find intervals $]\mu_j + \varepsilon_j, \mu_{j+1} - \delta_j[$ such that

$$H_B'(s) - H_B'(s_j^\alpha) \geq \tau > 0 \tag{A.9}$$

with

$$\tau = K \left(\frac{4t}{U}\right)^2 t \left(1 - \frac{4t}{U}\right)^{-1} \tag{A.10}$$

and $K$ the constant defined in section 3.2.

Explicit computation for the case $\mu \in ]\mu_3, \mu_{cb}[$ discussed in sec. 3.2, yields for $\cos \phi \geq 0$

$$\varepsilon_3 = 27 \cdot 2^{13} \left(1 - \frac{4t}{U}\right)^{-1} \frac{t^6}{U^5} \tag{A.11}$$

$$\delta_3 = 27 \cdot 2^{14} \left(1 - \frac{4t}{U}\right)^{-1} \frac{t^6}{U^5} \tag{A.12}$$

and the condition $\varepsilon_3 + \delta_4 < \mu_{cb} - \mu_3 = \frac{3}{2}(3 + \cos \phi)\frac{t^4}{U^3}$ will be satisfied if $U > 387$.

In the case of the bosons, as discussed in section 3.3, we can use the results for the fermions with constant external magnetic flux $\phi = \frac{\pi}{2}$.

## A.2  Triangular lattice

|  | $\alpha_1$ | $\alpha_2$ | $\alpha_3$ | $\alpha_4$ |
|---|---|---|---|---|
| $\mathcal{S}_1$, $\cos\phi \geq 0$ | $-\frac{1}{40} + \frac{\cos\phi}{120} + \frac{2\delta}{15}$ | $-\frac{1}{480} + \frac{\cos\phi}{40} + \frac{\delta}{15}$ | $\frac{1}{480} - \frac{\cos\phi}{240} - \frac{\delta}{240}$ | $-\frac{1}{32}$ |
| $\mathcal{S}_1$, $\cos\phi \leq 0$ | $-\frac{11}{240} - \frac{3\cos\phi}{40} + \frac{2\delta}{15}$ | $\frac{1}{120} + \frac{\cos\phi}{15} + \frac{\delta}{15}$ | $0$ | $-\frac{1}{24} - \frac{\cos\phi}{24}$ |
| $\mathcal{S}_2$, $\cos\phi \geq 0$ | $-\frac{1}{16} + \frac{\delta}{8}$ | $\frac{1}{24} + \frac{\cos\phi}{48} + \frac{\delta}{16}$ | $-\frac{1}{96} - \frac{\cos\phi}{32} + \frac{\delta}{96}$ | $-\frac{1}{16}$ |
| $\mathcal{S}_3$, $\cos\phi \geq 0$ | $\frac{\delta}{8}$ | $\frac{1}{96} + \frac{\cos\phi}{48} + \frac{\delta}{16}$ | $-\frac{1}{24} - \frac{\cos\phi}{32} + \frac{\delta}{96}$ | $-\frac{1}{32}$ |
| $\mathcal{S}_3$, $\cos\phi \leq 0$ | $\frac{\delta}{8}$ | $-\frac{5}{96} - \frac{\cos\phi}{96} + \frac{\delta}{12}$ | $-\frac{11}{480} - \frac{3\cos\phi}{80} + \frac{\delta}{240}$ | $-\frac{9}{160} - \frac{7\cos\phi}{160} + \frac{\delta}{120}$ |

Table 1: Coefficients for the definition of the zero potential $K_B$. This is for the truncated hamiltonian up to order 3 in the case of fermions on a square lattice. $\mu = -\frac{2t^2}{U} + \frac{t^4}{U^3}\delta$.

**Fermions**

The $3 \times 3$ blocks of the square lattice are replaced by hexagons $H$: the hexagon centered in $x$ is defined as $\{y \in \Lambda : |x - y| \leq 1\}$. The truncated hamiltonian may be written as

$$H^{(3)}(s;\mu) = \sum_\Delta \left( -\frac{1}{6}\mu \sum_{x \in \Delta} \frac{s_x + 1}{2} + \frac{t^2}{8U} \sum_{<x,y> \subset \Delta} s_x s_y \right.$$
$$\left. -\frac{t^3}{U^2}\frac{\cos\phi}{8} \sum_{x \in \Delta} s_x + \frac{3\cos\phi}{8}\frac{t^3}{U^2}s_\Delta \right) + \frac{t^4}{2U^3} \sum_{H \subset \Lambda} H_H(s) + \text{cte} \quad (A.13)$$

where

$$H_H(s) = -\frac{7 + 5\cos 2\phi}{32} \sum_{<x,y> \subset H} s_x s_y + \frac{4 - \cos 2\phi}{16} \sum_{\substack{(x,y) \subset H \\ |x-y|=\sqrt{3}}} s_x s_y$$
$$+ \frac{1}{4} \sum_{\substack{(x,y) \subset H \\ |x-y|=2}} s_x s_y + \frac{5\cos\phi}{16} \sum_{P \subset H} s_P \quad (A.14)$$

As before we introduce zero potentials $k_H^{(i)}$; numbering the sites starting from the center, i.e. 1 is the center and 2–7 are the external sites, we define

$$k_H^{(1)} = 6s_1 - \sum_{i=2}^{7} s_i \quad (A.15)$$

$$k_H^{(2)} = \sum_{i=2}^{7} s_1 s_i - \sum_{i=2}^{6} s_i s_{i+1} - s_2 s_7 \quad (A.16)$$

$$k_H^{(3)} = s_1(\sum_{i=2}^{5} s_i s_{i+2} + s_6 s_2 + s_7 s_3) - \sum_{i=2}^{5} s_i s_{i+1} s_{i+2} - s_6 s_7 s_2 - s_7 s_2 s_3 \quad (A.17)$$

We set as before $H'_H = H_H + K_H$ with $K_H$ a linear combination of $k_H^{(1)}$ and $k_H^{(2)}$ such that $H'_H(s;\mu)$ is an $m$-potential. Hereafter we define domains in the phase diagram, for which we can find an $m$-potential $K_H = \sum_i \alpha_i k_H^{(i)}$ with $\alpha_i$ given in Table 2. Unfortunately these domains do not cover the whole phase diagram, and thus there are still regions for which our results are not proven.

$\mathcal{D}_-$ : domain of $\mathcal{T}_-$

$\mathcal{D}_1$ : domain of $\mathcal{T}_1$

$\mathcal{D}_3$ : domain of $\mathcal{T}_3$

$\mathcal{D}_5^{(1)}$ : $\{(\mu,\phi); \tilde{\mu}_{cb} < \mu < -\frac{t^2}{U}\}$, with

$$\tilde{\mu}_{cb} = \begin{cases} -3\frac{t^2}{U} + 3\cos\phi\frac{t^3}{U^2} + (\frac{33}{4} + 9\cos 2\phi)\frac{t^4}{U^3} & \text{if } \cos 2\phi \geq 0 \\ -3\frac{t^2}{U} + 3\cos\phi\frac{t^3}{U^2} + (\frac{33}{4} + \frac{9}{2}\cos 2\phi)\frac{t^4}{U^3} & \text{if } \cos 2\phi < 0 \end{cases}$$

($\mathcal{D}_5^{(1)}$ is a part of the domain of $\mathcal{T}_5$)

$\mathcal{D}_5^{(2)}$ : $\{(\mu,\phi); \cos 2\phi \leq \frac{5}{12}$ and $-\frac{t^2}{U} < \mu < -6\cos\phi\frac{t^3}{U^2} + (-\frac{21}{4} + 9\cos 2\phi)\frac{t^4}{U^3}\}$

$\mathcal{D}_5^{(3)}$ : $\{(\mu,\phi); \cos 2\phi > \frac{5}{12}, -\frac{t^2}{U} < \mu < \mu_5$, and also :
$\mu < -6\cos\phi\frac{t^3}{U^2} + (-\frac{93}{128} + \frac{9\cos 2\phi}{16})\frac{t^4}{U^3}$ and $\mu < -6\cos\phi\frac{t^3}{U^2} + (\frac{27}{20} - \frac{9\cos 2\phi}{2})\frac{t^4}{U^3}\}$

($\mathcal{D}_5^{(2)}$ and $\mathcal{D}_5^{(3)}$ are also parts of the domain of $\mathcal{T}_5$)

$\mathcal{D}_6$ : domain of $\mathcal{T}_6$, under the restriction $\mu < -6\cos\phi\frac{t^3}{U^2} + (\frac{9}{4} - \frac{27\cos 2\phi}{2})\frac{t^4}{U^3}$

$\mathcal{D}_7$ : domain of $\mathcal{T}_7$, with $\mu < -6\cos\phi\frac{t^3}{U^2} + (\frac{9}{8} - \frac{9\cos 2\phi}{4})\frac{t^4}{U^3}$

$\mathcal{D}_8^{(1)}$ : domain of $\mathcal{T}_8$, with $\mu \geq -6\cos\phi\frac{t^3}{U^2} + (-\frac{3}{2} + 3\cos 2\phi)\frac{t^4}{U^3}$

$\mathcal{D}_8^{(2)}$ : domain of $\mathcal{T}_8$, with $\mu \leq -6\cos\phi\frac{t^3}{U^2} - \frac{9\cos 2\phi}{2}\frac{t^4}{U^3}$

$\mathcal{D}_9$ : domain of $\mathcal{T}_9$

| | $\alpha_1$ | $\alpha_2$ | $\alpha_3$ |
|---|---|---|---|
| $\mathcal{D}_-$ and $\mathcal{D}_1$ | $\frac{45}{224} - \frac{19\cos 2\phi}{224} - \frac{\delta'}{7}$ | 0 | 0 |
| $\mathcal{D}_3$ | $\frac{1}{8} - \frac{\cos 2\phi}{32} - \frac{\delta'}{8}$ | 0 | 0 |
| $\mathcal{D}_5^{(1)}$ | $\frac{1}{96} - \frac{5\cos 2\phi}{32} - \frac{\delta'}{9}$ | 0 | 0 |
| $\mathcal{D}_5^{(2)}$ | $\frac{3}{128} - \frac{3\cos 2\phi}{32} - \frac{\delta}{8}$ | $-\frac{11}{64} - \frac{9\cos 2\phi}{32} + \frac{\delta}{12}$ | $-\frac{17}{384} - \frac{\cos 2\phi}{32} + \frac{\delta}{72}$ |
| $\mathcal{D}_5^{(3)}$ | $\frac{1}{224} - \frac{3\cos 2\phi}{112} - \frac{5\delta}{42}$ | $-\frac{47}{224} - \frac{33\cos 2\phi}{224} + \frac{2\delta}{21}$ | $-\frac{17}{224} + \frac{9\cos 2\phi}{112} + \frac{\delta}{42}$ |
| $\mathcal{D}_6$ | $-\frac{\delta}{8}$ | $-\frac{33}{160} - \frac{87\cos 2\phi}{160} + \frac{\delta}{20}$ | $\frac{3}{80} - \frac{9\cos 2\phi}{40} + \frac{\delta}{40}$ |
| $\mathcal{D}_7$ | $\frac{1}{32} - \frac{\cos 2\phi}{16} - \frac{\delta}{9}$ | $-\frac{11}{32} + \frac{\cos 2\phi}{32} + \frac{\delta}{18}$ | $\frac{1}{32} - \frac{\cos 2\phi}{16} + \frac{\delta}{18}$ |
| $\mathcal{D}_8^{(2)}$ | $-\frac{\delta}{8}$ | $-\frac{9}{32} - \frac{3\cos 2\phi}{32} + \frac{\delta}{12}$ | $\frac{\delta}{24}$ |
| $\mathcal{D}_8^{(1)}$ and $\mathcal{D}_9$ | $-\frac{\delta}{8}$ | $-\frac{9}{32} - \frac{3\cos 2\phi}{32}$ | 0 |

Table 2: Coefficients for the $m$-potential. We defined $\delta$ s.t. $\mu = -6\cos\phi\frac{t^3}{U^2} + \delta\frac{t^4}{U^3}$ and $\delta'$ s.t. $\mu = -\frac{3t^2}{U} + \frac{3\cos\phi t^3}{U^2} + \frac{t^4}{U^3}\delta'$.

| | $\alpha_1$ | $\alpha_2$ | $\alpha_3$ |
|---|---|---|---|
| $\mu < \mu_2$ | $\frac{47}{112} - \frac{5\cos 2\phi}{28} - \frac{\delta}{7}$ | 0 | 0 |
| $\mu_2 < \mu < \mu_2$ | $\frac{11}{32} - \frac{\cos 2\phi}{8} - \frac{\delta}{8}$ | 0 | 0 |
| $\mu_3 < \mu < -\frac{t^2}{U}$ | $\frac{11}{48} - \frac{\cos 2\phi}{12} - \frac{\delta}{9}$ | 0 | 0 |
| $-\frac{t^2}{U} < \mu < \mu_4$ | $-\frac{\mu}{8}$ | $\frac{11}{16} - \frac{\cos 2\phi}{8} + \frac{\mu}{12}$ | $\frac{\cos 2\phi}{24} + \frac{\mu}{72}$ |
| $\mu_4 < \mu < (-2 - 3\cos 2\phi)\frac{t^4}{U^3}$ | $-\frac{\mu}{8}$ | $\frac{43}{80} - \frac{9\cos 2\phi}{40} + \frac{\mu}{20}$ | $\frac{1}{20} + \frac{3\cos 2\phi}{40} + \frac{\mu}{40}$ |
| $(-2 - 3\cos 2\phi)\frac{t^4}{U^3} < \mu \leq 0$ | $-\frac{\mu}{8}$ | $\frac{7}{16} - \frac{3\cos 2\phi}{8}$ | 0 |

Table 3: Coefficients for the $m$–potential of bosons on triangular lattice. We defined $\delta$ such that $\mu = -\frac{3t^2}{U} - \frac{3\cos\phi t^3}{U^2} + \frac{t^4}{U^3}\delta$. The $\mu_i, i = 2, 3, 4$ were previously defined.

**Bosons**

The construction of the appropriate $m$–potential is given in Table 3.

# B Proof of the theorem 4.1

We will show that for any configuration $\{\bar{\phi}_C + \delta_C\}$ with $\delta_C \neq 0$ for at least one cell $C$ we have

$$H_{eff}(s; \mu; \{\bar{\phi}_C + \delta_C\}) - H_{eff}(s; \mu; \{\bar{\phi}_C\}) > 0 \tag{B.1}$$

Recall that $H_{eff}$ is given by (2.16) where $\Omega(s)$ is the set of circuits $\omega$ which visit at least one empty and one occupied site. We introduce the set of circuits $\tilde{\Omega}(s)$ such that $\omega \in \Omega(s)$ and there exists a cell $C_\omega$ which is enclosed by $\omega$ but not by any shorter path. We remark that for any $\omega \in \tilde{\Omega}(s)$ the cell $C_\omega$ is unique and $N(\omega) = 1$ (Eq. (2.5)). We also denote by $k_C(\omega)$ the winding number of $\omega$ around the cell $C$. For $\omega \in \tilde{\Omega}(s)$, $k_C(\omega) = \pm 1$.

We split the left hand side of (B.1) in two parts $I_1 + I_2$, where $I_1$ comes from the circuits $\omega \in \tilde{\Omega}(s)$ and $I_2$ from the circuits $\omega \in \Omega(s) \setminus \tilde{\Omega}(s)$. For $I_1$ we have

$$I_1 = \sum_{\omega \in \tilde{\Omega}(s)} \frac{t^{|\omega|}}{U^{|\omega|-1}} g_\omega(s) \left[\cos\left\{\sum_C k_C(\omega)(\bar{\phi}_C + \delta_C)\right\}\right.$$
$$\left. - \cos\left\{\sum_C k_C(\omega)\bar{\phi}_C\right\}\right] \tag{B.2}$$

This term is decomposed further as $I_1 = J_1 + J_2$ with

$$J_1 = \sum_C \sum_{\substack{\omega \in \tilde{\Omega}(s) \\ C_\omega = C}} \frac{t^{|\omega|}}{U^{|\omega|-1}} g_\omega(s) \left[\cos\left\{\sum_{C'} k_{C'}(\omega)\bar{\phi}_{C'} + k_C(\omega)\delta_C\right\}\right.$$
$$\left. - \cos\left\{\sum_{C'} k_{C'}(\omega)\bar{\phi}_{C'}\right\}\right] \tag{B.3}$$

and

$$J_2 = \sum_C \sum_{\substack{\omega \in \tilde{\Omega}(s) \\ C_\omega = C}} \frac{t^{|\omega|}}{U^{|\omega|-1}} g_\omega(s) \left[ \cos \left\{ \sum_{C'} k_{C'}(\omega)(\bar{\phi}_{C'} + \delta_{C'}) \right\} \right.$$
$$\left. - \cos \left\{ \sum_{C'} k_{C'}(\omega)\bar{\phi}_{C'} + k_C(\omega)\delta_C \right\} \right]. \quad (B.4)$$

For $I_2$ we have the same formula than (B.2) with the sum over $\omega \in \tilde{\Omega}(s)$ replaced by the sum over $\omega \in \Omega(s) \setminus \tilde{\Omega}(s)$. In what follows we show that $J_1$ is bounded below by a strictly positive number and that $I_2$, $J_2$ are of smaller order.

**Lower bound for $J_1$**

Since $\sum_{C'} k_{C'}(\omega)\bar{\phi}_{C'} = 0$ or $\pi$ and $k_{C'}(\omega) = \pm 1$ the difference of the two cosines in $J_1$ is equal to

$$\cos \left\{ \sum_{C'} k_{C'}(\omega)\bar{\phi}_{C'} \right\} (\cos \delta_{C'} - 1) \quad (B.5)$$

Therefore

$$J_1 = \sum_C \Psi_C(s)(1 - \cos \delta_C) \quad (B.6)$$

with

$$\Psi_C(s) = \sum_{\substack{\omega \in \tilde{\Omega}(s) \\ C_\omega = C}} \frac{t^{|\omega|}}{U^{|\omega|-1}} |g_\omega(s)| (-1)^{m(\omega)-1} \cos \left\{ \sum_{C'} k_{C'}(\omega)\bar{\phi}_{C'} \right\} \quad (B.7)$$

One can then show that for any $\omega \in \tilde{\Omega}(s)$

$$(-1)^{m(\omega)-1} \cos \left\{ \sum_{C'} k_{C'}(\omega)\bar{\phi}_{C'} \right\} = 1 \quad (B.8)$$

and therefore $\Psi_C(s) > 0$. We give the details of the proof of (B.8) only for the square lattice, the case of the triangular lattice being similar.

$C_\omega$ is the unique cell surrounded by $\omega$, for which $\omega$ is a circuit of minimal length. Then necessarily $\omega$ is contained in a rectangle with $C_\omega$ in one corner, and the occupation of all sites in the rectangle is the same except for the site $x$ (or maybe the two sites $x$ and $y$) in the corner opposite to $C_\omega$ (fig. 17).

Therefore in the case of figure 17 (a), $\bar{\phi}_{C'} = 0$ for all $C'$ surrounded by $\omega$, and $(-1)^{m(\omega)} = -1$, which implies (B.8). On the other hand, in the case of figure 17 (b), $\bar{\phi}_{C'} = 0$ for all $C' \neq C_1$ surrounded by $\omega$ and $\bar{\phi}_{C_1} = \frac{1}{2}(1 + s_x s_y)\pi$. Thus if $\omega$ contains only one of the sites $(x, y)$ (B.8) follows as previously, while if $\omega$ contains both sites $x$ and $y$,

$$(-1)^{m(\omega)} \cos\left\{\sum_{C'} k_{C'}(\omega)\bar{\phi}_{C'}\right\} = s_x s_y \cos\left\{\frac{1}{2}(1 + s_x s_y)\pi\right\} = -1.$$

Figure 17: Minimal circuit for an empty square; $s_x$ and/or $s_y = -s$.

Since $m$ is either 1 or $j - 1$, $j - 2$, we have in all cases

$$\Psi_C(s) \geq \frac{1}{2^{j(C)-1}} \frac{t^{j(C)}}{U^{j(C)-1}} \tag{B.9}$$

where $j(C)$ is the length of the shortest path $\omega \in \tilde{\Omega}(s)$ surrounding $C$. Since $(1 - \cos \delta_C) \geq \frac{2}{\pi^2}\delta_C^2$ we get from (B.6)

$$J_1 \geq \sum_C \frac{1}{\pi^2} \frac{1}{2^{j(C)-2}} \frac{t^{j(C)}}{U^{j(C)-1}} \delta_C^2 \tag{B.10}$$

Now we proceed to find upper bounds for $I_2$ and $J_2$.

**Upper bound for $I_2$**

Using $|g_\omega(s)| \leq \frac{1}{2}$ for all $\omega$, $|\sin \alpha| \leq |\alpha|$, and $\cos \alpha - \cos \beta = -2 \sin \frac{\alpha+\beta}{2} \sin \frac{\alpha-\beta}{2}$, we may write

$$|I_2| \leq \sum_{\omega \in \Omega(s)\setminus\tilde{\Omega}(s)} \frac{1}{2}\frac{t^{|\omega|}}{U^{|\omega|-1}} 2 \left|\frac{1}{2}\sum_C k_C(\omega)\delta_C\right|^2$$

$$\leq \frac{1}{4} \sum_{\omega \in \Omega(s)\setminus\tilde{\Omega}(s)} \frac{t^{|\omega|}}{U^{|\omega|-1}} \left[\sum_C |k_C(\omega)|\right] \sum_C |k_C(\omega)|\delta_C^2 \tag{B.11}$$

where the second inequality comes from Cauchy-Schwartz. We bound $\sum_C |k_C(\omega)|$ by $\gamma |\omega|^2$, with $\gamma$ a constant depending on the lattice (square lattice : $\gamma = \frac{1}{16}$; triangular lattice : $\gamma = \frac{1}{6}$). Then

$$|I_2| \leq \frac{\gamma}{4} \sum_C \delta_C^2 \sum_{\omega \in \Omega(s)\setminus\tilde{\Omega}(s)} |\omega|^2 \frac{t^{|\omega|}}{U^{|\omega|-1}} |k_C(\omega)| \tag{B.12}$$

We have $|k_C(\omega)| \leq \frac{|\omega|}{|C|}$. Moreover since the circuits belong to $\Omega(s)\backslash\tilde{\Omega}(s)$ and surround $C$ their minimum length is $j(C) + 1$; therefore the number of circuits of length $n$ enclosing $C$ is less than $\gamma' n^2 z^{n-1}$, with the maximum coordination number $z$ and the constant $\gamma'$ depending only on the lattice. Thus we obtain

$$\begin{aligned}|I_2| &\leq \frac{\gamma\gamma'}{4|C|}\sum_C \delta_C^2 \sum_{n\geq j(C)+1} n^5 \frac{t^n}{U^{n-1}} z^{n-1} \\ &\leq \frac{\gamma\gamma'}{4|C|}\sum_C \delta_C^2 \frac{t^{j(C)}}{U^{j(C)-1}} z^{j(C)-1} \sum_{n\geq 1}(n+j(C))^5 \left(\frac{t}{U}\right)^n z^n\end{aligned} \quad (B.13)$$

with the hypothesis of theorem 4.1 there exists $j_{\max}$ such that $j(C) \leq j_{\max}$ and therefore

$$|I_2| \leq a\frac{t}{U}\sum_C \frac{t^{j(C)}}{U^{j(C)-1}}\delta_C^2 \quad (B.14)$$

where the constant $a$ does not depend on $\frac{t}{U}$.

**Upper bound for $J_2$**

Proceeding as in the previous paragraph we write

$$|J_2| \leq \sum_{\omega\in\tilde{\Omega}(s)} \frac{1}{2}\frac{t^{|\omega|}}{U^{|\omega|-1}} 2 \left|\frac{1}{2}\sum_{C\neq C_\omega} k_C(\omega)\delta_C\right|\left|\frac{1}{2}\sum_{C\neq C_\omega}k_C(\omega)\delta_C + k_{C_\omega}(\omega)\delta_{C_\omega}\right|$$

$$\leq \frac{\gamma}{4}\sum_C \delta_C^2 \sum_{\substack{\omega\in\tilde{\Omega}(s)\\C_\omega\neq C}} \frac{t^{|\omega|}}{U^{|\omega|-1}}|\omega|^2|k_C(\omega)| + \frac{1}{2}\sum_C|\delta_C|\sum_{\substack{\omega\in\tilde{\Omega}(s)\\C_\omega\neq C}}\frac{t^{|\omega|}}{U^{|\omega|-1}}|k_C(\omega)||\delta_{C_\omega}| \quad (B.15)$$

The first sum may be bounded as previously and

$$|J_2| \leq a\frac{t}{U}\sum_C \delta_C^2 \frac{t^{j(C)}}{U^{j(C)-1}} + \frac{1}{2}\sum_C\sum_{C'\neq C}|\delta_C||\delta_{C'}|\sum_{\substack{\omega\in\tilde{\Omega}(s)\\C_\omega=C'}}\frac{t^{|\omega|}}{U^{|\omega|-1}}|k_C(\omega)| \quad (B.16)$$

We remark for later use that in the last sum the lengths of circuits $\omega$ are equal to $j(C')$, and since they surround $C$ we have $j(C) < j(C')$.

**Proof of (B.1)**

Collecting the three bounds (B.10), (B.14) and (B.16) we get

$$\begin{aligned}H_{eff}(s;\mu;\{\bar{\phi}_C+\delta_C\}) &- H_{eff}(s;\mu;\{\bar{\phi}_C\}) = J_1 + J_2 + I_2 \\ &\geq \sum_C b(C)\frac{t^{j(C)}}{U^{j(C)-1}}\delta_C^2 - \frac{1}{2}\sum_{\substack{C,C'\\C\neq C'}}|\delta_C||\delta_{C'}|\sum_{\substack{\omega\in\tilde{\Omega}(s)\\C_\omega=C'}}\frac{t^{j(C')}}{U^{j(C')-1}}|k_C(\omega)|\end{aligned} \quad (B.17)$$

where

$$b(C) = \frac{1}{\pi^2}\frac{1}{2^{j(C)-2}} - 2a\frac{t}{U}$$

Let $K(C)$ be the number of occurences of $C$ in the second sum; then $K(C) \leq \gamma'' j_{\max}^2$, with $\gamma''$ depending only on the lattice, and

$$\begin{aligned}
&H_{eff}(s;\mu;\{\bar{\phi}_C + \delta_C\}) - H_{eff}(s;\mu;\{\bar{\phi}_C\}) \\
&\geq \sum_C \frac{b(C)}{K(C)+1} \frac{t^{j(C)}}{U^{j(C)-1}} \delta_C^2 + \sum_{\substack{C,C' \\ j(C)<j(C')}} \left\{ \frac{b(C)}{K(C)+1} \frac{t^{j(C)}}{U^{j(C)-1}} \delta_C^2 \right. \\
&\quad + \frac{b(C)}{K(C')+1} \frac{t^{j(C')}}{U^{j(C')-1}} \delta_{C'}^2 - \frac{1}{2}|\delta_C||\delta_{C'}| \sum_{\substack{\omega \in \tilde{\Omega}(s) \\ C_\omega = C'}} \frac{t^{j(C')}}{U^{j(C')-1}} |k_C(\omega)| \right\} \\
&\geq \sum_C \frac{b(C)}{K(C)+1} \frac{t^{j(C)}}{U^{j(C)-1}} \delta_C^2 \\
&\quad + \sum_{\substack{C,C' \\ j(C)<j(C')}} \left\{ \sqrt{\frac{b(C)}{K(C)+1}} \frac{t^{\frac{j(C)}{2}}}{U^{\frac{j(C)-1}{2}}} |\delta_C| - \sqrt{\frac{b(C)}{K(C')+1}} \frac{t^{\frac{j(C')}{2}}}{U^{\frac{j(C')-1}{2}}} |\delta_{C'}| \right\}^2 \quad (B.18)
\end{aligned}$$

In writing this last inequality we used

$$\frac{1}{2} \sum_{\substack{\omega \in \tilde{\Omega}(s) \\ C_\omega = C'}} \frac{t^{j(C')}}{U^{j(C')-1}} |k_C(\omega)| \leq \frac{2\sqrt{b(C)b(C')}}{\sqrt{(K(C)+1)(K(C')+1)}} \frac{t^{\frac{j(C)+j(C')}{2}}}{U^{\frac{j(C)+j(C')}{2}-1}} \quad (B.19)$$

which is true for $U$ large enough, since $j(C) < j(C')$. Theorem 4.1 follows from (B.18).

# C  Flux phase problem

## C.1  Square lattice

From (2.15) we obtain an expression similar to (A.2) with

$$H_B(s;\mu) = -\delta(s_{x(B)}+1) + \frac{1}{12} \sum_{\substack{(x,y) \subset B \\ |x-y|=2}} s_x s_y$$

$$+ \frac{1}{32} \sum_{P \subset B} \left[ \left(1 - \sum_{(x,y) \subset P} s_x s_y + 5 s_P \right) \cos \bar{\phi}_P - \frac{7}{2} \sum_{<x,y> \subset P} s_x s_y + 4 \sum_{\substack{(x,y) \subset P \\ |x-y|=\sqrt{2}}} s_x s_y + 2 \right] \quad (C.1)$$

As before we consider the new potential $H'_B = H_B + \sum_{i=1}^{4} \alpha_i k_B^{(i)}$ with the coefficients $\alpha_i$ defined in Table 4 and the zero–potentials $k_B^{(i)}$ given by Eq. (A.5)–(A.8). One can check that $H'_B$ is an $m$–potential leading to the phase diagram of fig.15.

|     | $\alpha_1$ | $\alpha_2$ | $\alpha_3$ | $\alpha_4$ |
| --- | --- | --- | --- | --- |
| $\mathcal{S}_1$ | $\frac{\delta}{8}$ | $-\frac{1}{16}+\frac{\delta}{16}$ | $\frac{7}{32}+\frac{\delta}{96}$ | $-\frac{1}{32}$ |
| $\mathcal{S}_3$ | $-\frac{3}{80}+\frac{2\delta}{15}$ | $-\frac{29}{480}+\frac{\delta}{15}$ | $\frac{1}{4}$ | $-\frac{1}{24}$ |

Table 4: Zero–potentials $K_B$ for the fermions on a square lattice, with minimal flux.

## C.2 Triangular lattice

When $\mu$ is close to $-3\frac{t^2}{U}-3\frac{t^3}{U^2}$ the ground state configurations consist of triangles containing zero or one ion; otherwise the energy increases as $\frac{t^2}{U}$.

The minimum flux for these configurations is $\pi$ everywhere; since it is constant we can use the results of section 3.4 and thus the domains of $\mathcal{T}_-$, $\mathcal{T}_1$, $\mathcal{T}_3$ and part of $\mathcal{T}_5$ are proven.

For the remaining part we write the truncated hamiltonian as

$$H^{(3)}(s;\mu) = \sum_{\Delta\subset\Lambda}\left(\frac{t^2}{8U}\sum_{<x,y>\subset\Delta}s_xs_y+\frac{t^3}{8U^2}\left[3s_\Delta-\cos\bar\phi_\Delta\sum_{x\in\Delta}s_x\right]\right)$$
$$+\frac{t^4}{2U^3}\sum_{H\subset\Lambda}H_H(s;\mu) \quad (C.2)$$

with

$$H_H(s;\mu) = -\frac{7}{32}\sum_{<x,y>\subset H}s_xs_y+\frac{1}{4}\sum_{\substack{(x,y)\subset H\\|x-y|=\sqrt{3}}}s_xs_y+\frac{1}{4}\sum_{\substack{(x,y)\subset H\\|x-y|=2}}s_xs_y$$
$$+\frac{1}{16}\sum_{P\subset H}\cos\bar\phi_P\left(1-\sum_{(x,y)\subset P}s_xs_y+5s_P\right) \quad (C.3)$$

where $\bar\phi_P = \sum_{\Delta\subset P}\bar\phi_\Delta$.

Adding to $H_H(s;\mu)$ the zero–potentials defined by (A.15)–(A.17) with the factors given in Table 5, we obtain an $m$–potential proving the phase diagram of fig.16.

|     | $\alpha_1$ | $\alpha_2$ | $\alpha_3$ |
| --- | --- | --- | --- |
| $\mathcal{T}_5$ | $-\frac{\mu}{8}$ | $-\frac{7}{32}+\frac{\mu}{12}$ | $\frac{1}{24}+\frac{\mu}{72}$ |
| $\mathcal{T}_6$ | $-\frac{\mu}{8}$ | $-\frac{15}{32}+\frac{\mu}{20}$ | $\frac{1}{8}+\frac{\mu}{40}$ |
| $\mathcal{T}_9$ | $-\frac{\mu}{8}$ | $-\frac{23}{32}$ | $0$ |

Table 5: Zero–potentials $K_B$ for the fermions on a triangular lattice, with minimal flux.